1

# Efficient Surface Formation Route of Interstellar Hydroxylamine through NO Hydrogenation II: the multilayer regime in interstellar relevant ices

G. Fedoseev[1, a)], S. Ioppolo[1, ‡], T. Lamberts[1, 2], J. F. Zhen[1], H. M. Cuppen[2], H. Linnartz[1]

[1]*Sackler Laboratory for Astrophysics, Leiden Observatory, University of Leiden, PO Box 9513, NL 2300 RA Leiden, The Netherlands*

[2]*Institute for Molecules and Materials, Radboud University Nijmegen, PO Box 9010, NL 6500 GL Nijmegen, The Netherlands*

**Hydroxylamine ($NH_2OH$) is one of the potential precursors of complex pre-biotic species in space. Here we present a detailed experimental study of hydroxylamine formation through nitric oxide (NO) surface hydrogenation for astronomically relevant conditions. The aim of this work is to investigate hydroxylamine formation efficiencies in polar (water-rich) and non-polar (carbon monoxide-rich) interstellar ice analogues. A complex reaction network involving both final ($N_2O$, $NH_2OH$) and intermediate (HNO, $NH_2O·$, etc.) products is discussed. The main conclusion is that hydroxylamine formation takes place via a fast and barrierless mechanism and it is found to be even more abundantly formed in a water-rich environment at lower temperatures. In parallel, we experimentally verify the non-formation of hydroxylamine upon UV photolysis of NO ice at cryogenic temperatures as well as the non-detection of NC- and NCO-bond bearing species after UV processing of NO in carbon monoxide-rich ices. Our results are implemented into an astrochemical reaction model, which shows that $NH_2OH$ is abundant in the solid phase under dark molecular cloud conditions. Once $NH_2OH$ desorbs from the ice grains, it becomes available to form more complex species (*e.g.*, glycine and β -alanine) in gas phase reaction schemes.**

## I. INTRODUCTION

The observed chemical complexity in space results from the cumulative outcome of gas, grain and gas-grain interactions. Recent studies indicate that a substantial part of the stable and complex species identified so far, form on icy dust grains.[1, 2] Interstellar grains, indeed, act as micrometer-sized cryo-pumps providing surfaces onto which gas-phase species can accrete, meet and react. During the last decade several independent laboratory studies showed the relevance of such astrochemical solid-state reactions. Meanwhile, the surface formation of the bulk of the identified interstellar ices (*i.e.*, water, methanol, carbon dioxide, formaldehyde, and formic acid) has been confirmed through subsequent H-atom addition reactions to simple species, like CO- and/or $O_2$-ices.[3-10]

In space, non-energetic atom addition induced solid state chemistry occurs mostly at low temperatures (~10 K), *i.e.*, in the innermost part of circumstellar clouds where newly formed molecules are, to a great extent, shielded from radiation by dust particles. These regions are part of collapsing envelopes feeding new stars - young stellar objects (YSOs) - and provide the original material from which comets and ultimately planets are made. [11] Since solid-state neutral-neutral reactions often do not require activation energy, [1, 2] they play a key role in the chemistry of dark interstellar clouds. In the final stages of star formation energetic processing - like UV photolysis, cosmic ray irradiation and thermal processing - also contributes to surface reaction schemes by adding external energy into the ice. It has indeed been demonstrated that complex organic molecules and even amino acids are formed after energetic processing of interstellar ice

---

a) Author to whom correspondence should be addressed. Electronic mail: fedoseev@strw.leidenuniv.nl

‡) New Address: Division of Geological and Planetary Science, California Institute of Technology, 1200 E. California Blvd., Pasadena, California 91125, USA

analogues.[12, 13] Therefore, laboratory based studies of the solid-state formation of organic material is of considerable interest, since efficient reaction routes provide a recipe to form pre-biotic species in star and planet forming regions.[14]

Hydroxylamine ($NH_2OH$) is an important precursor species in the formation of amino acids.[15] Several formation schemes of hydroxylamine have been proposed in the past. Blagojevic *et al.*[15] describe $NH_2OH$ formation through UV photolysis of $NH_3 + H_2O$ and/or "NO + 3H" ices. Charnley *et al.*[14] suggested that hydroxylamine is formed through non-energetic hydrogenation of NO ice under dark cloud conditions. Nitric oxide has been detected in the gas phase towards many dark and warm clouds with relative abundances of $1·10^{-8}$ - $1·10^{-7}$ with respect to $H_2$.[16-19] Detailed astrochemical models[14, 19, 20] indicate that the main formation route for NO in the gas-phase is a neutral-neutral reaction $N + OH \rightarrow NO + H$, while the main gas-phase destruction channel is $NO + N \rightarrow N_2 + O$. Under dense cloud conditions, gas-phase NO can accrete, like CO, on the surface of dust grains and is, therefore, expected to participate in the solid-state chemical network leading to the formation of N- and NO-bearing species. Moreover, solid NO may also be formed through surface reactions in the early stages of quiescent dark clouds.

Although NO has the potential to be one of the main precursors of complex molecules in space, solid state reaction schemes involving NO-bearing ices have not yet been studied in detailed experiments. Recently, we presented for the first time hydroxylamine formation via a non-energetic solid-phase route - pure NO + H on crystalline $H_2O$ ice as well as on other different substrates[21] – and we showed the astrochemical impact of that scheme. In the present paper, the focus is on the investigation of the underlying physical-chemical processes that lead to surface hydroxylamine formation in the multilayer regime. Thus, experiments are extended to more realistic and interstellar relevant polar ($NO:H_2O + H/D$) and non-polar (NO:CO + H/D) environments. In the following sections we investigate the effect of temperature, ice composition and H-atom flux on the $NH_2OH$ and HNO formation efficiencies. We also discuss experiments in which NO, $NO:H_2O$ and NO:CO ices are UV irradiated with Ly-α light. The astrophysical implications of this work are briefly discussed. In an accompanying and complementary paper[22] (published back-to-back with the present one) NO hydrogenation reactions are studied with the focus on the submonolayer regime using interstellar relevant substrates, *i.e.*, amorphous silicate and crystalline $H_2O$ ice.

## II. EXPERIMENTAL DETAILS AND DATA REDUCTION

### A. Experimental procedure

The experiments are performed using two similar ultra-high vacuum (UHV) setups: a SURFace REaction SImulation DEvice (SURFRESIDE), optimized to study H-atom addition reactions, and a CRYOgenic Photoproduct Analysis Device (CRYOPAD), designed for UV photolysis experiments of interstellar ice analogues.

#### 1. SURFRESIDE

Thermal H/D-atom addition reactions are investigated using SURFRESIDE which consists of a stainless steel UHV main chamber and an atomic line. Experimental details are available in Refs. 4, 6, 9. A rotatable gold-coated copper substrate is placed in the centre of an UHV chamber. The room temperature base pressure in the main chamber is $<3.5·10^{-10}$ mbar. The substrate temperature is controlled from 12 to 300 K using a He closed-cycle cryostat. Ice deposition proceeds under an angle of 45º, with controllable rates from 0.5 to 2.5 Langmuir per minute (L/min, where 1 L = $1.3·10^{-6}$ mbar s$^{-1}$). A quadrupole mass spectrometer (QMS) placed behind the substrate is used to monitor the main-chamber gas composition. The atomic line faces the sample and comprises a well-characterized hydrogen thermal cracking source[23] used to produce non-energetic H/D atoms from $H_2/D_2$ molecules. Hydrogen/deuterium molecules are cracked by passing through a capillary pipe surrounded by a heated tungsten filament. Dissociation of molecules occurs through collisions with the hot (1850 ºC) walls of the capillary pipe with a degree of dissociation >35%.[23] A nose-shaped quartz pipe is placed along the path of the beam and is used to thermalize both H/D atoms and non-dissociated $H_2/D_2$ molecules to room temperature before they reach the surface of the ice sample. H/D-atom fluxes are measured quantitatively by placing the QMS at the substrate position and are chosen between $7·10^{12}$ and $3·10^{13}$ atoms cm$^{-2}$ s$^{-1}$. This procedure has been described in detail in Ref. 24.

A high vacuum glass manifold with a base pressure $<5·10^{-5}$ mbar is used for gas mixture preparation. To avoid water contamination the glass line is connected to a liquid nitrogen trap and is flushed well by mixture components. A new mixture is prepared for each experiment. After preparation, the mixture is introduced to a pre-pumped stainless steel dosing line ($<1·10^{-5}$ mbar). The metal line is typically filled with a gas pressure of 30 mbar and is kept isolated for the duration of the experiment. The deposition rate is controlled by a precise all-metal leak valve. The gases used in this work are: NO (Linde 99.5 %), CO (Linde 99.997%), $N_2O$ (Praxair 99.5%), $NO_2$ (Linde 99%) and a Milli-Q water sample degassed under high vacuum conditions.

The ice composition is monitored *in situ* by means of Reflection Absorption InfraRed Spectroscopy (RAIRS) in the range: 700 – 4000 cm$^{-1}$ (14 – 2.5 μm) with a spectral resolution of 0.5 cm$^{-1}$ using a Fourier transform infrared spectrometer. Two different experimental procedures are applied here. In "pre-deposition" experiments, ices are first deposited on a gold surface and are subsequently exposed to a thermal H/D-atom beam. In this case RAIR difference spectra are acquired during hydrogenation/deuteration of the sample with respect to a background spectrum of the initial deposited ice at low temperature. In the case of "co-deposition" experiments, nitrogen oxide bearing ices are continuously deposited simultaneously with H/D atoms. The formation of intermediate species and final products is monitored in the ice by changing ice deposition rates and H/D-atom fluxes, as discussed in Ref. 25. In this case RAIR difference spectra are acquired during co-deposition with respect to a background spectrum of the bare gold substrate at a low temperature.

At the end of every H/D-atom exposure a temperature programmed desorption (TPD) experiment is performed. Because the QMS is situated behind the substrate, RAIR spectra and QMS data cannot be taken simultaneously during the heating phase, and crucial experiments are performed twice in order to interpret TPD results using both RAIRS and QMS techniques. In the first experiment, the sample is rotated 180º to face the QMS and it is heated linearly at a rate of 1 K/min. Desorbed species are subsequently detected as a function of sample temperature using the QMS. In the second experiment, the sample is not rotated and RAIR spectra of the ice are acquired every 5 K while heating at a rate of 1 K/min. In this way infrared spectroscopic and mass spectrometric information can be combined. It should be noted that a TPD experiment implicitly leads to the morphological modification and eventual destruction of the ice. Therefore TPD data are mainly used in our experiments as an additional tool to constrain the low temperature RAIR results.

*2. CRYOPAD*

UV photolysis experiments are performed using CRYOPAD which is described in detail in Ref. 26 and references therein. Similarly to SURFRESIDE, the rotatable gold-coated substrate in CRYOPAD is connected to a He close-cycle cryostat and placed in the centre of a stainless steel UHV chamber. The room temperature base pressure of the system is <$2.5 \cdot 10^{-9}$ mbar and the temperature of the substrate is controlled between 15 and 200 K. Ice deposition proceeds at an angle of 90º and a rate of 1.5 L/min. After deposition, the ice film is irradiated by UV light from a broadband hydrogen microwave discharge, which peaks at 121 nm (Ly-α) and covers the range from 115 – 170 nm (7 – 10.5 eV) with an incidence angle of 45º. Photolysis products upon UV irradiation are monitored by means of the same FTIR spectrometer that is used for SURFRESIDE, covering 700 – 4000 $cm^{-1}$ with 0.5 $cm^{-1}$ resolution. A QMS is incorporated into the setup in order to monitor molecules in the gas phase and specifically photodesorption products. At the end of a UV irradiation experiment, a TPD experiment is routinely performed. The setup is constructed such that the QMS already faces the sample, and consequently, during a TPD it is possible to monitor ice constituents both spectroscopically and mass spectrometrically. The heating rate used for these experiments is the same as for SURFRESIDE (1 K/min). The UV flux is measured indirectly by calibration via the previously studied photodesorption rate of a pure CO ice: carbon monoxide ice is exposed to Ly-α light and the number of photodesorbed CO molecules is determined using the RAIR difference spectra. The UV flux is then derived, assuming that the CO photodesorption rate obtained here and previously in Ref. 27 are identical. Typical UV fluxes amount to $1.5 \cdot 10^{14}$ photons $cm^{-2}$ $s^{-1}$.

Gas mixtures used during experiments with CRYOPAD are prepared separately in the same glass manifold as described above. The mixtures are then introduced into a pre-pumped stainless steel dosing line (<$1 \cdot 10^{-4}$ mbar) that is always filled twice in order to avoid the decomposition of gas mixture components on the metal walls during the preparation hours: a first time (at the beginning of the experiment) to select at room temperature the desired deposition rate by adjusting a precise all-metal leak valve and monitoring the increase in pressure in the main chamber; and a second time (right before deposition) using the same pressure used the first time to fill the dosing line. When not in use, the deposition line is kept under high vacuum. In this way highly reproducible deposition rate ca be guaranteed.

The experiments at SURFRESIDE and CRYOPAD focus on selected surface reactions involving the hydrogenation and UV irradiation of pure NO ices and mixtures of NO with CO, $H_2O$ and $N_2$ over a wide range of laboratory conditions including different atomic fluxes, ice temperatures, co-deposition rates and mixture ratios. A number of control experiments are performed to prove that newly detected species are formed upon H/D-atom exposure or UV photolysis on top of or in the ice sample, and that they are not the result of contamination or background gas-phase reactions. All experiments are summarized in Table I. Moreover, in order to unambiguously identify absorption bands of possible reaction products we have performed separate experiments to acquire spectra of pure $N_2O$ as well as $NO:N_2O_3$ and $NO_2:N_2O_4$ ice mixtures (see Table II) to make direct comparisons possible.

Table I. List of performed experiments.

| Depositing ice composition | Type | Temperature (K) | Deposition rate (L/min) | H/D flux (atom·cm$^{-2}$·s$^{-1}$) | H$_2$/D$_2$ flux (mol·cm$^{-2}$·s$^{-1}$) | UV flux (ph·cm$^{-2}$·s$^{-1}$) | Time (min) |
|---|---|---|---|---|---|---|---|
| Co-deposition:[a] | | | | | | | |
| NO | +H | 15 | 0.5 | 3·10$^{13}$ | 3.5·10$^{15}$ | - | 200 |
| -//- | +H$_2$ | 15 | 0.5 | 0 | 3.5·10$^{15}$ | - | 40 |
| -//- | +H | 15 | 2.5 | 7·10$^{12}$ | 2.5·10$^{14}$ | - | 120 |
| -//- | +H | 25 | 2.5 | 7·10$^{12}$ | 2.5·10$^{14}$ | - | 100 |
| -//- | +D | 15 | 0.5 | 3·10$^{13}$ | 3.5·10$^{15}$ | - | 120 |
| -//- | +D | 15 | 2.5 | 7·10$^{12}$ | 2.5·10$^{14}$ | - | 120 |
| NO:N$_2$ (1:5) | +H | 15 | 0.5 | 3·10$^{13}$ | 3.5·10$^{15}$ | - | 100 |
| -//- | +H$_2$ | 15 | 2.5 | 0 | 3.5·10$^{15}$ | - | 30 |
| NO:CO (1:1) | +H | 15 | 0.5 | 3·10$^{13}$ | 3.5·10$^{15}$ | - | 200 |
| -//- | +H | 15 | 2.5 | 7·10$^{12}$ | 2.5·10$^{14}$ | - | 120 |
| -//- | +D | 15 | 0.5 | 3·10$^{13}$ | 3.5·10$^{15}$ | - | 200 |
| -//- | +D | 15 | 2.5 | 7·10$^{12}$ | 2.5·10$^{14}$ | - | 120 |
| NO:CO (1:6) | +H | 15 | 0.5 | 3·10$^{13}$ | 3.5·10$^{15}$ | - | 420 |
| -//- | +H | 15 | 2.5 | 3·10$^{13}$ | 3.5·10$^{15}$ | - | 120 |
| -//- | +H | 15 | 2.5 | 7·10$^{12}$ | 2.5·10$^{14}$ | - | 120 |
| -//- | +H | 25 | 0.5 | 3·10$^{13}$ | 3.5·10$^{15}$ | - | 120 |
| -//- | +D | 15 | 2.5 | 3·10$^{13}$ | 3.5·10$^{15}$ | - | 120 |
| NO:H$_2$O (1:6) | +H | 15 | 2.5 | 7·10$^{12}$ | 2.5·10$^{14}$ | - | 150 |
| -//- | +H | 15 | 2.5 | 3·10$^{13}$ | 3.5·10$^{15}$ | - | 60 |
| Pre-deposition:[b] | | | | | | | |
| NO | +H | 15 | 2.5 (50 L) | 3·10$^{13}$ | 3.5·10$^{15}$ | | 150 |
| NO:CO:N$_2$ (1:1:5) | +H | 15 | 2.5 (100 L) | 3·10$^{13}$ | 3.5·10$^{15}$ | | 60 |
| Nitrogen oxides deposition:[c] | | | | | | | |
| N$_2$O | | 30 | 2.5 (100 L) | - | - | - | - |
| N$_2$O | +H | 15 | 0.5 | 3·10$^{13}$ | 3.5·10$^{15}$ | - | 120 |
| NO$_2$[d] | | 15 | 2.5 (30 L) | - | - | - | - |
| NO$_2$[d] | +H | 15 | 2.5 | 7·10$^{12}$ | 2.5·10$^{14}$ | - | 40 |
| NO$_2$:NO (1:2)[e] | | 15 | 2.5 (60 L) | - | - | -- | - |
| Pre-deposition (UV-photolysis): | | | | | | | |
| NO | +hυ | 15 | 1.0 (40 L) | - | | 1.5·10$^{14}$ | 130 |
| NO:CO (1:1) | +hυ | 15 | 1.0 (40 L) | - | - | 1.5·10$^{14}$ | 75 |
| NO:CO (1:6) | +hυ | 15 | 1.0 (40 L) | - | - | 1.5·10$^{14}$ | 150 |
| NO:H$_2$O (1:6) | +hυ | 15 | 1.0 (40 L) | - | - | 1.5·10$^{14}$ | 120 |
| NO:H$_2$O:CO (1:1:1) | +hυ | 15 | 2.5 (80 L) | - | - | 1.5·10$^{14}$ | 160 |
| NO:H$_2$O:CO (1:6:6) | +hυ | 15 | 2.5 (80 L) | - | - | 1.5·10$^{14}$ | 140 |
| NO:CH$_4$ (1:1) | +hυ | 15 | 2.5 (80 L) | - | - | 1.5·10$^{14}$ | 120 |

[a]Co-deposition is an experiment in which NO containing ices are deposited during H/D flux exposure.
[b]Pre-deposition is an experiment in which NO containing ices are first deposited and subsequently exposed to H/D atom or UV fluxes.
[c]This is a set of experiments in which the spectra of different nitrogen oxides were obtained and then monitored for stability upon H-atom flux exposure.
[d]NO$_2$ gas contained considerable NO admixture; formation of N$_2$O$_3$ (ON-NO$_2$) along with N$_2$O$_4$ (O$_2$N-NO$_2$) is observed upon deposition.
[e]Mixture of N$_2$O$_3$ (ON-NO$_2$) and cis-(NO)$_2$ is formed upon deposition.

**B. Data Analysis**

Straight baseline segments are subtracted from all acquired spectra. Subsequently, the bands areas corresponding to species present in the ice are integrated. As previously discussed in Refs. 4, 6 transmission band strengths cannot be used directly to derive column densities of species in reflection experiments. Moreover, isothermal desorption experiments, which provide us with absorbance per monolayer, cannot be performed for the unstable intermediates that are detected during co-deposition experiments. Therefore, we refrain from deriving column densities and instead follow the formation trends of the detected species only by integrating the corresponding band areas as a function of time (see Results and Discussion). As a consequence, a full quantitative characterization is not possible. However, we can compare formation trends of the same species from different experiments and thereby derive information on temperature dependence, ice composition effects and the reaction network. All the assigned absorption features that correspond to nitrogen-bearing species and present in our experiments are summarized in Table II. The asterisk in Table II marks the spectral features used for integration and relative quantification.

Table II. List of assigned nitrogen-bearing species.

| Mode | Frequency (cm$^{-1}$) | Regular species | Frequency (cm$^{-1}$) | Deuterated species | Reference |
|---|---|---|---|---|---|
| $\nu_5$ [a] | 1776 | cis-$(NO)_2$ | | | 28, 29 |
| $\nu_1$ | 1865 | cis-$(NO)_2$ | | | 28, 29 |
| | 1741 | trans-$(NO)_2$ | | | 28 |
| | 1875 | NO(monomer) | | | 28, 29 |
| $\nu_6$ | 889 | $NH_2OH$ | 825 | $ND_2OD$ | 30, 31 |
| $\nu_6$ | 919 | $NH_2OH$ (bulk) | 878 | $ND_2OD$ (bulk) | 32 |
| $\nu_5$ [a] | 1144 | $NH_2OH$ | 920 | $ND_2OD$ | 30, 31 |
| $\nu_5$ [a] | 1203 | $NH_2OH$ (bulk) | 946 | $ND_2OD$ (bulk) | 32 |
| $\nu_4$ | 1359 | $NH_2OH$ | 1026 | $ND_2OD$ | 30, 31 |
| $\nu_4$ | 1514 | $NH_2OH$ (bulk) | 1126 | $ND_2OD$ (bulk) | 32 |
| $\nu_3$ | 1592 | $NH_2OH$ | 1175 | $ND_2OD$ | 30, 31 |
| $\nu_3$ | 1608 | $NH_2OH$ (bulk) | 1185 | $ND_2OD$ (bulk) | 32 |
| | 2716 | $NH_2OH$ (bulk) | 2045 | $ND_2OD$ (bulk) | 32 |
| $\nu_2$ | 2899 | $NH_2OH$ (bulk) | 2184 | $ND_2OD$ (bulk) | 32 |
| $\nu_2$ | 3194 | $NH_2OH$ (bulk) | 2393 | $ND_2OD$ (bulk) | 32 |
| $\nu_7$ | 3261 | $NH_2OH$ (bulk) | | | 32 |
| $\nu_1$ | 3317 | $NH_2OH$ (bulk) | 2482 | $ND_2OD$ (bulk) | 32 |
| $\nu_3$ | 1286 | $N_2O$ | | | 33 |
| $\nu_1$ [a] | 2235 | $N_2O$ | | | 33 |
| $\nu_2$ | 1507 | HNO | 1156 | DNO | 34 |
| $\nu_3$ [a] | 1561 | HNO | 1546 | DNO | 34 |
| | 1829 | X-NO | 1823 | X-NO | - |
| $\nu_1$ | 1851 | $N_2O_3$ | | | 28, 35 |
| $\nu_2$ | 1614 | $N_2O_3$ | | | 28, 35 |
| $\nu_3$ | 1302 | $N_2O_3$ | | | 28, 35 |
| $\nu_4$ | 782 | $N_2O_3$ | | | 28, 35 |

[a]The spectral features used for integration and relative quantification

All absorbance features chosen for data analysis have a good signal-to-noise ratio and do not overlap with other absorbance features. The range of integration is set manually for each species and kept the same for all the spectra acquired during a single experiment. The error bars that are indicated in the figures listed in the Results and Discussion section are estimated as follows: several blank spectra, *i.e.*, without deposited ice but with a cold substrate, are acquired before each experiment. The corresponding baseline segments are subtracted from the blank data. The noise area in the blank spectra is integrated over the same frequency range that is used for band integration of the selected species. The uncertainties are then derived by averaging the values of the measured blank area for each selected range. Although these error bars do not include deviations in the baseline subtraction procedure, they provide lower limits for the detectable signal.

## III. RESULTS AND DISCUSSION

### A. Hydrogenation of pure NO ice

Figure 1(a) shows RAIR difference spectra recorded with SURFRESIDE of a pre-deposited NO ice after exposure to a thermal H-atom beam at 15 K. The negative peaks around 1776 and 1865 cm$^{-1}$ correspond to *cis*-(NO)$_2$ [28, 29] used up during hydrogenation of the ice. Due to the small exothermicity of NO dimerization (14 kJ per mol)[35] and a fast radical-radical recombination reaction rate, nitric oxide forms solid *cis*-(NO)$_2$ immediately after deposition. Therefore, NO monomers are detected in the ice by means of RAIRS only when NO is trapped in an ice matrix of other species (see below). Thus, hydrogenation/deuteration of a pure NO ice involves as a first step the hydrogenation/deuteration of (NO)$_2$ dimers. The positive peaks in Fig. 1(a) indicate the presence of newly formed species in the ice. However, their identification is not clear *a priori* because of their low signal to noise ratio. A possible reason for this low coverage of processed ice can be found in the structure of the ice. Crystalline (NO)$_2$ has a compact monoclinic unit cell with a space group P2$_1$/a,[36] which is more comparable to a CO than an O$_2$ ice structure.[37, 38] Although our deposited ice is not necessarily crystalline, the local structure of (NO)$_2$ agglomerates is probably close to crystalline. As shown in Refs. 4, 9 a compact ice structure prevents H/D atoms from penetrating deep into the bulk of the ice. As a consequence, in a pre-deposition experiment only the first few monolayers are actively involved in surface reactions, the final yield of H/D atom addition products is low and no intermediate products are detected.

Figures 1(b) and (c) show NO/H-atom co-deposition experiments at 15 K. The main advantage of co-deposition experiments is the ability to change the NO/H(D) ratio in order to: 1) trap non-fully hydrogenated products inside the growing ice by using a high NO flow with a low H(D)-atom flux (Fig. 1(c) and (e)); or 2) deposit a fully hydrogenated thick ice by using a low NO flow with a high H(D)-atom flux (Fig. 1(b) and (d)). Growing a thick hydrogenated ice during co-deposition experiments increases the column densities of the reaction products and allows their unambiguous identification *in situ* and at low temperatures by means of RAIRS. Although QMS is more sensitive than RAIRS, it allows for the detection of species only above their desorption temperatures. For the species studied here, this is higher than 15 K, *i.e.*, >75 K for N$_2$O and >140 K for NH$_2$OH.

Several positive absorption features, *e.g.*, at 1608 cm$^{-1}$ (NH$_2$ scissoring), 1514 cm$^{-1}$ (HON bending), 1203 cm$^{-1}$ (NH$_2$ wagging), and 919 cm$^{-1}$ (ON stretching),[31] present in Fig. 1(b) clearly show that the main hydrogenation product of a pure NO ice at low temperature is solid hydroxylamine. The NH$_2$OH formed during the co-deposition experiment shown in Fig 1(b) is bulk hydroxylamine, bound with four different types of hydrogen bonds, such as OH⋯H, OH⋯N, NH⋯O, NH⋯N, and with different strengths. The hydroxylamine OH and NH stretching modes are in the range from 2600 to 3400 cm$^{-1}$ and overlap with each other. There are two groups of broad bands in this region, at 3317, 3261 and 3194 cm$^{-1}$ and at 2899 and 2716 cm$^{-1}$. The first group can be assigned to the symmetric and asymmetric N-H stretches, and the O-H···O stretch, while the second group is due to a hydrogen bonded OH···N stretch.[31, 39] The formation of hydroxylamine in our pre-deposition experiments is confirmed by comparing all the hydroxylamine peaks present in Fig. 1(b) with those shown in Fig. 1(a).

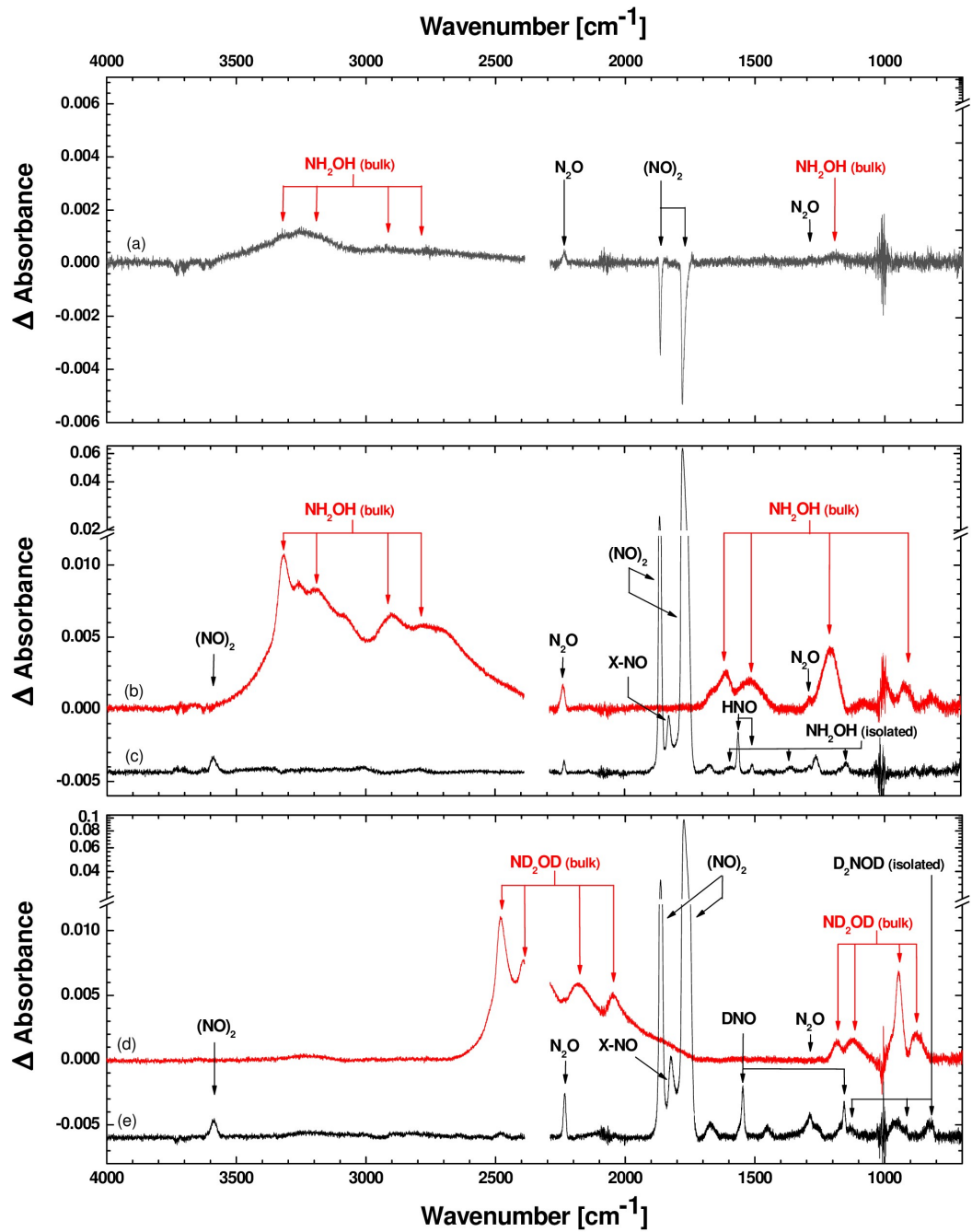

Figure 1. RAIR difference spectra of H/D atom addition to a pure NO ice for a sample temperature of 15 K. The upper panel (a) shows a spectrum of a pre-deposited NO (50 L) after exposure to an H-atom beam. The middle panel shows co-deposition of NO and H atoms with NO:H = 1:10 (b) and 2:1 (c). In the lower panel two spectra recorded upon co-deposition of NO and D atoms with NO:D = 1:10 (d) and 2:1 (e) are presented.

Figure 1(c) presents several features due to the $(NO)_2$ dimer, $N_2O$, and isolated $NH_2OH$, and reveals several new features: two narrow bands at 1561 and 1507 cm$^{-1}$ that can be assigned to HNO, and another two at 1829 cm$^{-1}$ and 1680 cm$^{-1}$ with an unclear origin. The feature at 1829 cm$^{-1}$ overlaps with the NO stretching mode which may indicate that this species contains weakly bound NO molecules. In the following sections we will refer to this unknown component as X-NO. A broad peak at 1680 cm$^{-1}$ is present in almost every spectrum and demonstrates similar behaviour under various experimental conditions, thus interpretation of this feature is uncertain. To verify our assignments we repeated the same co-deposition experiments discussed above using D atoms instead of H atoms (Figs. 1(d) and (e)).

TPD experiments further constrain the formation of $NH_2OH$ and $N_2O$. Under our experimental conditions the desorption of $NH_2OH$ starts at 140 K and peaks at 175 K (TPD rate 1 K/min). The $NH_2OH$ desorption peak is reported at 188 K in Ref. 22. There a desorption energy of 54.2 kJ mol$^{-1}$ is calculated for pure $NH_2OH$ from a surface of amorphous silicate. The difference in desorption temperature between the present work and Ref. 22 is not surprising, as we study ices in the multilayer regime. Here desorption of $NH_2OH$ takes place from the surface of the ice itself. Therefore, under this regime molecules from a layer of $NH_2OH$ are weakly bound to other molecules (like $NH_2OH$, or $NH_2OH$ mixed with CO and/or $H_2O$, see sections below). This leads to a low temperature desorption peak. In Ref. 22, at most one monolayer of $NH_2OH$ is deposited onto an amorphous silicate surface. In that case $NH_2OH$ molecules are bound to the silicate surface and not to another layer of ice. The resulting binding energy is higher, leading to a higher temperature desorption peak.

**B. Hydrogenation of NO in non-polar CO surroundings**

We performed a series of experiments aimed at studying the effect of NO dilution in CO and $N_2$ (non-polar) matrices on the hydrogenation pathways. These experiments simulate the formation of hydroxylamine under dense and cold interstellar cloud conditions, when gas-phase CO freezes out. Figure 2 shows several examples of RAIR spectra of NO:CO co-deposition experiments with different ice mixtures (1:1, 1:6) and H/D-atom ratios.

All bands observed in a pure NO co-deposition experiment (Figs. 1(b) and (c)) are present here, and in addition, new features appear. In Figure 2(a) the NO monomer feature peaks at 1875 cm$^{-1}$,[28, 29] while the *trans*-$(NO)_2$ absorbance band is centred around 1745 cm$^{-1}$.[28] The presence of the NO monomer in the ice mixture indicates that the NO mobility is limited in a CO lattice at 15 K. The remaining absorption features belong to the products of H/D-atom addition to CO, such as $H_2CO$ (1720, 1499 cm$^{-1}$) and $CH_3OH$ (1060 cm$^{-1}$). These reactions have been examined extensively in previous studies.[3, 4] In the case of D-atom exposure traces of $D_2CO$ are detected at 1674 and 2092 cm$^{-1}$ but deuterated methanol is not found. It should be noted that formaldehyde is detected only in the experiments where all NO is converted into its final hydrogenation products (Figs. 2(b) and (d)). In the experiments where nitric oxide is still present in the ice, no products of H/D-atom addition to CO ice are found. This observation will be addressed in the following sections. Furthermore, no absorption features that can be assigned to N-C bond bearing species, like ·NCO radicals, HNCO or $NH_2CHO$, are found by RAIRS or QMS. Unstable intermediates like HCO and HNOH are not detected either. Note that due to the abundance of the $^{13}CO$ isotope in the CO gas cylinder (1,1 %), a clear $^{13}CO$ absorbance feature at 2092 cm$^{-1}$ is also observed**.**

Unlike CO ice, $N_2$ ice is inert to H-atom addition. The hydrogenation of nitric oxide in a nitrogen matrix with a ratio of NO:$N_2$ = 1:5 shows qualitatively that the final products from NO + H are the same as in a pure NO hydrogenation co-deposition experiment (Fig. 1(c)) with the difference that this time a small feature around 1875 cm$^{-1}$ can be assigned to the NO monomer (as in Fig. 2). Here $N_2O$ seems to be formed more efficiently compared to the NO hydrogenation experiment in a CO matrix.

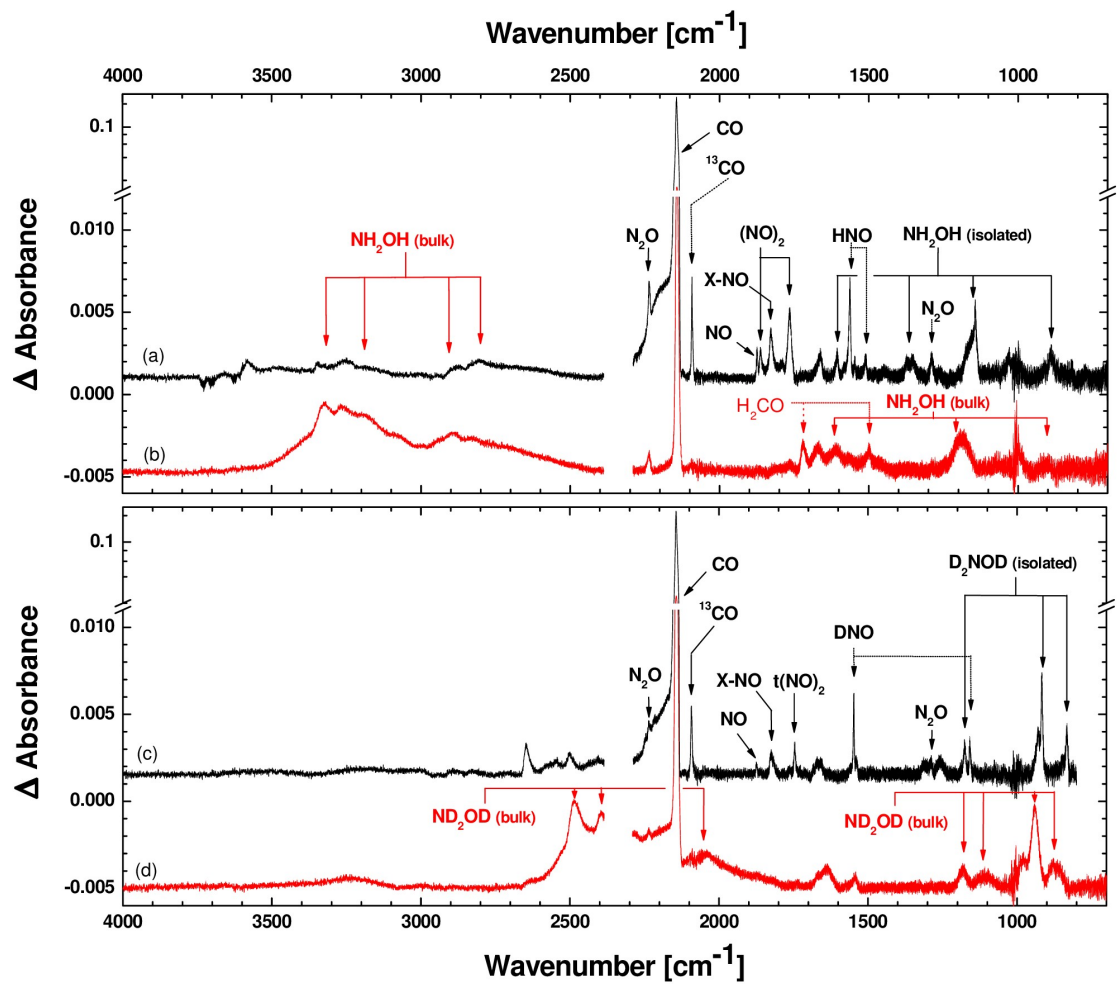

Figure 2. RAIR difference spectra of H/D atom additions to NO in non-polar CO ice for a sample temperature of 15 K. The upper panel shows co-deposition spectra of NO, CO and H atoms with NO:CO:H = 1:6:15 (a) and 1:1:20 (b). The lower panel shows two spectra for the co-deposition of NO, CO and D atoms with a ratio of 1:6:15 (c) and 1:1:20 (d).

## C. Polar $H_2O$ surroundings

Water is the most abundant species detected in interstellar ices towards dark clouds, low-mass and high-mass YSOs. It is generally accepted that water ice is formed mainly during the early stages of a translucent and quiescent cloud.[1, 14] Gas-grain chemical models predict that the NO abundance peaks later during CO freeze out.[14] Solid hydroxylamine can be formed during the same early stages in star forming regions from accreted NO or from NO formed on the surface of grains. Therefore, we studied the hydrogenation of NO in a water (polar) matrix (NO: $H_2O$ = 1:6). $H_2O$ molecules are inert to H-atom addition, but provide hydrogen bonds that may interact with NO molecules as well as possible reaction products upon NO hydrogenation, *i.e.*, HNO and $NH_2OH$. Here, an NO:$H_2O$ (1:6) ratio is used for comparison with the aforementioned non-polar mixtures.

Water ice deposited at 15 K forms an amorphous bulk with several very broad bands: the OH stretching mode between 3000 and 3700 cm$^{-1}$, the HOH bending mode in the 1400 – 1700 cm$^{-1}$ range and the libration mode between 700 and 1000 cm$^{-1}$. All these bands are present in our spectra (see Fig. 3). As opposed to NO:CO ices, no nitric oxide monomer and *trans*-(NO)$_2$ are detected in the ice, which can be explained by a different mobility of NO in water ice compared to NO in a CO matrix. Although most of the NH$_2$OH absorption features overlap with the water OH stretching and HOH bending modes, all of them can be found in Fig. 3 (b), where the H-atom flux is high enough to efficiently convert all NO to hydroxylamine. Figure 3 (a) shows a RAIR spectrum after co-deposition of NO:H$_2$O with a low H-atom flux. Here unlike for Fig. 3 (b), HNO, X-NO and N$_2$O features can be observed in the ice.

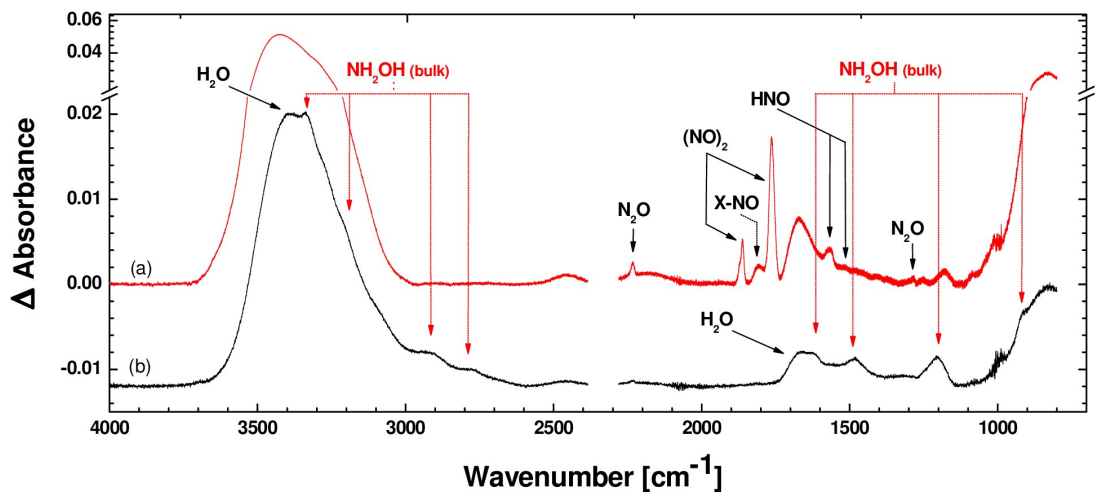

Figure 3. RAIR difference spectra of H-atom additions to NO in a polar H$_2$O ice for a sample temperature of 15 K. Co-deposition results of NO and H$_2$O with H atoms for NO:H$_2$O:H = 1:6:3.5 (a) and 1:6:15 (b) are shown.

**D. UV photolysis of NO containing ices**

Several pre-deposition experiments were performed with CRYOPAD to study the UV photon induced chemistry in pure NO ice and in ice mixtures containing nitric oxide in water or carbon monoxide. In addition, tertiary ice mixtures NO:CO:H$_2$O with ratios 1:1:1 and 1:6:6 were exposed to UV photons.

As discussed in the previous sections, pure NO converts efficiently into *cis*-(NO)$_2$ ice upon deposition (Fig. 4(b)). Several positive bands appear after UV processing of *cis*-(NO)$_2$ ice: the features at 2235 and 1286 cm$^{-1}$ already observed in the NO hydrogenation experiments are assigned to N$_2$O. New bands with maxima at 1602, 1299, 779 cm$^{-1}$ and, although not well resolved, 1850 cm$^{-1}$ appear only in the UV photolysis experiments. These can be assigned to N$_2$O$_3$ (see Fig. 4(a)).[28, 35, 40] The dilution of NO molecules in a CO matrix does not seem to influence this outcome substantially. As for Figure 4(a), all the absorption features assigned to *cis*-(NO)$_2$, N$_2$O and N$_2$O$_3$ are present in the RAIR spectra after exposure of NO:CO = 1:6 ice to UV photons (see Fig. 4(c)). In addition, the NO monomer is clearly detected because of the low mobility of NO in a CO lattice before and after UV photolysis of the ice sample. Moreover, the NO monomer abundance increases upon UV irradiation.

Two more carbon bearing species, apart from $^{12}$CO and $^{13}$CO are detected in an NO:CO ice upon UV photolysis: carbon dioxide and its $^{13}$C isotope analogue. Since the solid carbon dioxide IR absorption feature overlaps with atmospheric gas-phase CO$_2$ present outside the UHV chamber and along the line of the FTIR beam, the formation of carbon dioxide in the ice upon UV photolysis is constrained by QMS data.

Figures 4(e-f) show the resulting spectra upon UV photolysis of an NO:H$_2$O (1:6) ice mixture. Also in this case, two species are present in the ice upon deposition: *cis*-(NO)$_2$ and H$_2$O. UV processing of the sample induces formation of several products: N$_2$O and monomeric NO, hydrogen peroxide (H$_2$O$_2$) and traces of HNO (right shoulder at 1565 cm$^{-1}$). Absorption features at 1607 and 1303 cm$^{-1}$ may be due to N$_2$O$_3$ or NO$_2$. The absence of an absorption feature at 1850 cm$^{-1}$ makes the NO$_2$ assignment more likely,

but due to the low final yield this is not conclusive.

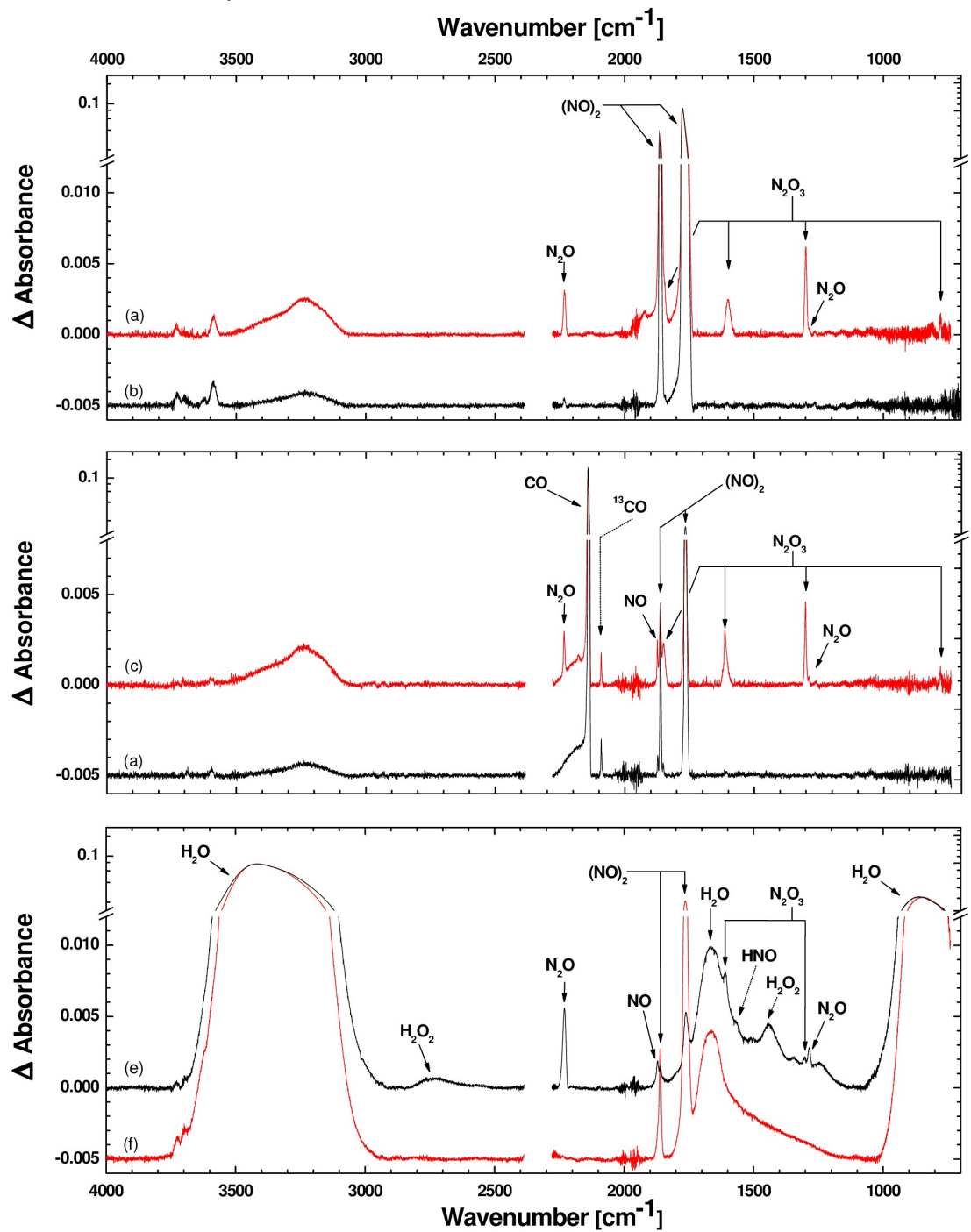

Figure 4. RAIR difference spectra obtained after UV photolysis of pre-deposited NO containing ices for a sample temperature of 15 K. The upper panel shows spectra of pure NO ice after deposition (b) and after exposure to $1\cdot10^{18}$ photons cm$^{-2}$ (a). The middle panel shows similar spectra of NO in non-polar CO ice for NO:CO = 1:6 before (d) and after exposure to $1\cdot10^{18}$ photons cm$^{-2}$ (c). The lower panel shows spectra of NO in polar H$_2$O ice after exposure to a fluence of $1\cdot10^{18}$ photons cm$^{-2}$ (e) and after deposition (f) for comparison.

The UV processing of the tertiary NO:CO:H$_2$O ice mixtures reveals the formation of the same species formed in the aforementioned NO:CO and NO:H$_2$O ice photolysis experiments. Here the same features are assigned to H$_2$CO with peaks at 1715 and 1500 cm$^{-1}$ and tentatively assigned to the NO stretching mode of HNO$_2$ peaking at 1640 cm$^{-1}$. It should be noted that the final yield of solid CO$_2$ is higher than in the case of UV irradiated NO:CO experiments. The addition of water ice to the NO:CO ice mixture increases the carbon dioxide formation. This can be explained by the presence of free OH radicals in the ice after UV irradiation that can react with CO to form CO$_2$ (more in the 'UV processing' section).

As can be concluded from Fig. 4, NCO radicals, NC$^-$ and NCO$^-$ anions as well as their hydrogenated analogues HNC and HNCO[41-44] are not present in any of our RAIR spectra. Another main result here is that NH$_2$OH is not formed upon UV photolysis of water rich ices. The straightforward conclusion is that surface H-atom addition reactions are needed to provide a pathway for the formation of NH$_2$OH in simple interstellar ice analogues. We will address these points in the following sections.

**E. Temperature dependence**

Temperature effects on reaction mechanisms can be rather complex and reflect the cumulative outcome of different temperature dependent processes. First of all, a higher ice temperature means more thermal energy that can help to overcome activation barriers. This has a direct effect on the final yields of the formed species and, therefore, on the formation trends. Secondly, a higher temperature can have an effect on the structure of the ice leading to molecular reorganization. For instance, it has been shown that different deposition temperatures result in different ice structures, *e.g.*, amorphous water ice versus crystalline water ice.[45, 46] This affects the penetration depth of H atoms in the ice and more generally diffusion processes. Then, the hopping and swapping rates of molecules in the ice increase with temperature, while the residence lifetime of a species on the ice surface decreases.

The temperature dependence of selected surface formation reaction pathways is studied here by repeating the same experiments for two different and astronomically relevant temperatures, *i.e.*, 15 and 25 K. Since the thermal desorption temperatures of CO and N$_2$ are around this value, higher temperatures have not been studied. The top panel of Fig. 5 shows the formation trends of HNO and NH$_2$OH at 15 and 25 K as a function of the H-atom fluence for a co-deposition experiment of pure NO and H atoms. Here the integrated absorption of the formed species presents a linear trend because of the constant co-deposition of the parent species. The bottom left-side panel of Fig. 5 shows NH$_2$OH, HNO and N$_2$O final yields from the aforementioned pure NO + H experiments, while the bottom right-side panel shows NH$_2$OH, HNO and N$_2$O final yields for two NO:CO/H co-deposition experiments at 15 and 25 K. All these experiments present lower final product yields at higher temperatures. Moreover, as shown in both bottom panels, the ratios between HNO and NH$_2$OH final yields at 15 and 25 K are roughly constant and are 1.5 and 3.5 for the pure NO + H and the NO:CO + H experiments, respectively. This indicates that the formation of HNO and NH$_2$OH are linked and are most likely temperature independent, *i.e.*, reactions take place without an activation barrier. The lower final yields at higher temperature can be explained by the lifetime of H atoms on the ice surface that, for instance, is 10$^3$ times shorter at 25 K than at 15 K on water ice.[47] In this way H atoms have less time to react with NO molecules leading to a decrease in the hydrogenation final yields. All final yields in the bottom right panel are lower than their counterparts in the bottom left panel. This can be explained by a lower abundance of NO molecules in the ice and the presence of solid CO. For the pure NO + H experiments, the final yield of solid N$_2$O is the same at different temperatures within the experimental uncertainties, while it decreases for higher temperatures in the NO:CO + H experiments. This indicates that the formation of N$_2$O requires a temperature dependent reaction scheme and a more complex ice composition.

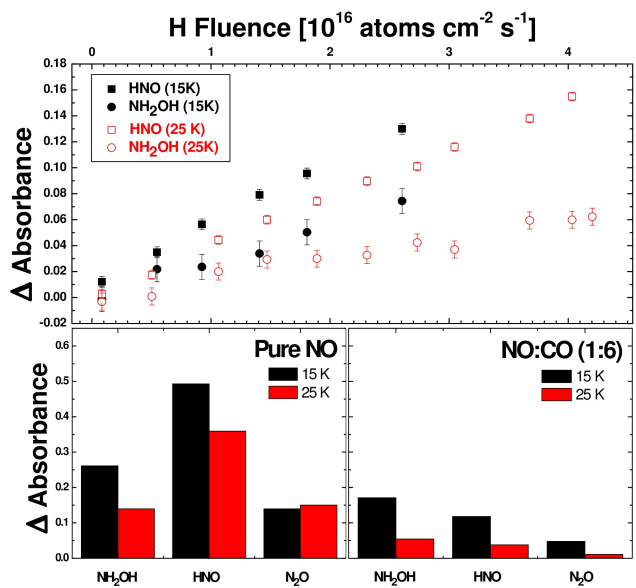

Figure 5. The top panel shows the temperature dependence of HNO and NH$_2$OH abundances versus the H-atom fluence for NO:H (2:1) co-deposition experiments at 15 and 25 K. The lower panels show the final yields of N$_2$O, HNO and NH$_2$OH for an NO:H (2:1) co-deposition experiment (bottom left) and an NO:CO:H (1:6:15) co-deposition experiment (bottom right). The left panel final yields are extrapolated from experimental data at the same H-atom fluence used for the final yields shown in the right panel.

**F. Matrix effects**

The dilution of NO in polar (water-rich ice) and non-polar (water-poor ice) matrices has a number of important effects on the hydrogenation pathways: 1) NO hydrogenation products can react with components of the mixture other than hydrogen atoms leading to a more complex reaction scheme; 2) the other molecules present in the deposited matrix (like CO) can participate in H-atom addition reactions, effectively influencing reaction rates through additional H-atom consumption. Their reaction products may interact with NO and NO hydrogenation reaction products; 3) other species in the ice can trap NO molecules in the bulk of the ice hiding them from the impinging H atoms; 4) H-bonding can improve coupling and heat dissipation through the ice, promoting the stabilization of intermediate products and changing reaction branching ratios.

In order to investigate non-polar ices we have compared the results from the hydrogenation of pure NO ice with those from the hydrogenation of NO:CO ice mixtures with ratios of 1:1 and 1:6. The choice of these ratios is not random: if NO hydrogenation products interact with those from CO hydrogenation, then the 1:1 ratio should provide the highest yield of the final products; on the other hand, for an NO:CO = 1:6 ratio a single NO molecule is surrounded on average by an octahedron of six CO molecules (the NO dimer is surrounded by 12 CO molecules). Therefore, this is a good approach to see whether NO and CO interact with each other upon hydrogenation of the ice. Moreover, since CO is highly abundant in quiescent dark clouds compared to NO,[16-19] a 1:6 ratio is astrochemically more relevant. We furthermore investigated the two NO:CO 1:1 and 1:6 mixtures for two different H-atom fluxes of $7 \cdot 10^{12}$ and $3 \cdot 10^{13}$ atom·cm$^{-2}$·s$^{-1}$. During these co-deposition experiments the ratio between H atoms and the sum of NO and CO molecules is kept constant. Figure 6 plots the integrated intensities of NO hydrogenation products as a function of the H-atom fluence for pure NO and NO:CO mixtures with ratios 1:1 and 1:6 co-deposited with H atoms.

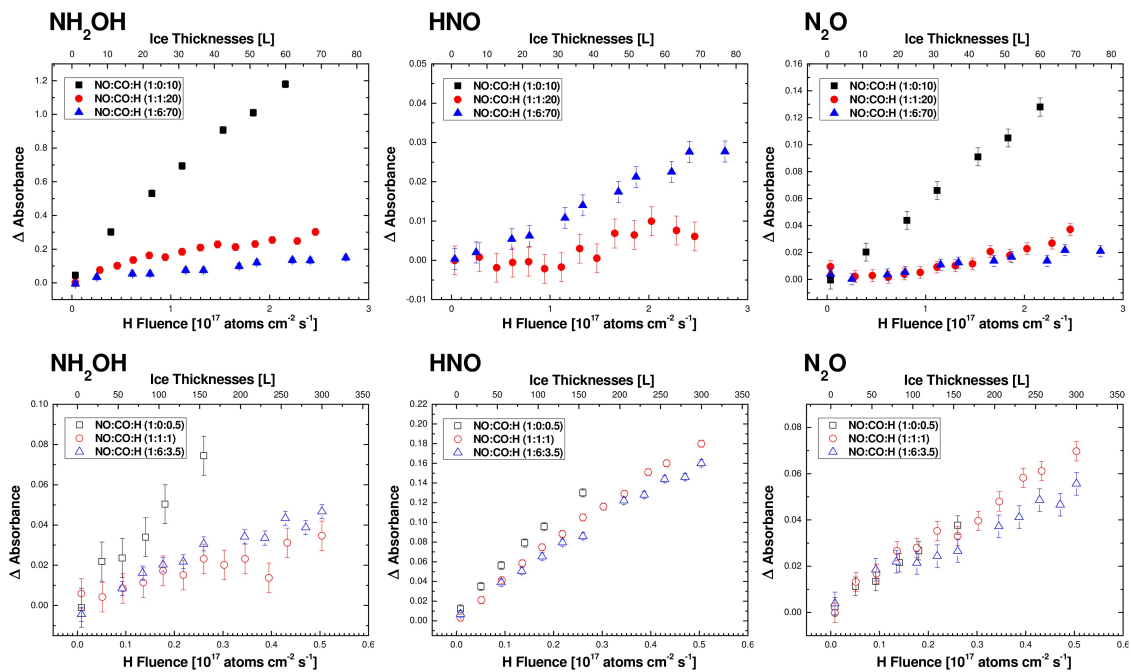

Figure 6. RAIR integrated intensities of hydroxylamine (left-side panels), nitroxyl (centre panels) and nitrous oxide (right-side panels) versus H-atom fluence for several co-deposition experiments using high H-atom flux (top panels) and low H-atom flux (bottom panels). Every panel shows three curves: one for H atoms co-deposited with pure NO (black squares), another for H atoms co-deposited with NO:CO = 1:1 mixture (red circles) and one for co-deposition of H atoms co-deposited with NO:CO = 1:6 mixture (blue triangles). A missing curve indicates that the product is not detected during the experiment.

The upper panels of Fig. 6 show the integrated intensities of $NH_2OH$, HNO and $N_2O$ versus the H-atom fluence for three co-deposition experiments (NO/H = 1/10, NO:CO/H = 1:1/20, 1:6/70) where the H-atom flux is high in order to efficiently convert nitric and carbon oxides into their hydrogenation products. The abundance of $NH_2OH$ in the ice is proportional to the abundance of NO in the mixture, while the HNO abundance is inversely proportional to that of NO. The $N_2O$ ice abundance is strongly affected by the presence of CO in the ice, however it does not seem to depend on the mixture ratios used here. In addition, and as mentioned before, NO absorption features (from both the dimer and monomer) are not present in the RAIR spectra of pure NO + H and NO:CO = 1:1 + H (see Fig. 1(b) and Fig. 2(b)) but only in the NO:CO = 1:6 hydrogenation experiment (Fig. 2(a)). The presence of NO in the ice after hydrogenation can be explained by the trapping effect of NO inside a CO lattice and the consequent blocking effect of H atoms in a CO ice.[4, 9] Consequently, the gradual decrease of the hydroxylamine yield with increasing CO content in the ice can be explained by a decrease in the absolute amount of NO available for hydrogenation reactions. This conclusion is further constrained by the nitroxyl final yield, which is inversely proportional to the hydroxylamine yield: HNO is only observed in the ice when some NO remains unconverted into $NH_2OH$. This confirms that the formation paths of nitroxyl and hydroxylamine are linked.

The lower panels of Fig. 6 show the RAIR integrated absorbances of $NH_2OH$, HNO and $N_2O$ versus the H-atom fluence for three co-deposition experiments (NO/H = 1/0.5, NO:CO/H = 1:1/1, 1:6/3.5) but now the H-atom flux is low in order to detect the not-fully-hydrogenated species in the ice. All plots show very similar abundances of $N_2O$ and HNO and slightly higher yields for $NH_2OH$ in a pure NO hydrogenation experiment compared to the hydrogenation of NO:CO mixtures. The products of H-atom addition to CO molecules ($H_2CO$, $CH_3OH$) are not detected in any of these experiments. This leads to an important conclusion: the hydrogenation of NO ice is more efficient than the hydrogenation of CO ice and is most likely barrierless as concluded here and in our previous work.[21] It was shown before that the intermediate CO + H and $H_2CO$ + H reaction steps have an activation barrier, while HCO + H and $H_3CO$ + H take place without an activation barrier.[3, 4] Thus, if the subsequent hydrogenation of nitric oxide (NO + H, HNO + H, $H_2NO$ + H) proceeds with no activation barrier, then the $NH_2OH$ final yield depends only on the amount of NO molecules and H atoms available for reaction. This is fully consistent with the observation that all infrared $NH_2OH$, HNO and $N_2O$ integrated band intensities shown in the lower panels of Fig. 6 present a similar final yield. Minor differences among the formation trends may be due to the presence of CO molecules.

The dipole moments of CO, NO and the *cis*-$(NO)_2$ are all small and comparable,[48-50] and the ice structures are also quite similar.[36, 38] This explains why dilution of NO in a CO matrix does not substantially influence the reaction network and, consequently, the formation trends of the final products. In contrast, water molecules have a strong dipole moment and form strong hydrogen bonds in the ice with other species, such as $NH_2OH$, $NH_2O\cdot$ and $NHOH\cdot$ radicals, and $H_2O$ itself. We performed several experiments co-depositing $NO:H_2O$ binary mixtures and H atoms using similar experimental conditions to those adopted for the non-polar ice investigation. The results are summarized in Fig. 7. In this case we compared results for an $NO:H_2O$ = 1:6 ratio co-deposition experiment with low and high H-atom fluxes. For a high H-atom flux the water matrix significantly increases the hydroxylamine final yield, while it decreases both nitroxyl and nitrous oxide final yields. For low H-atom flux only hydroxylamine exhibits a higher yield in a water lattice compared to a similar experiment with NO:CO = 1:6.

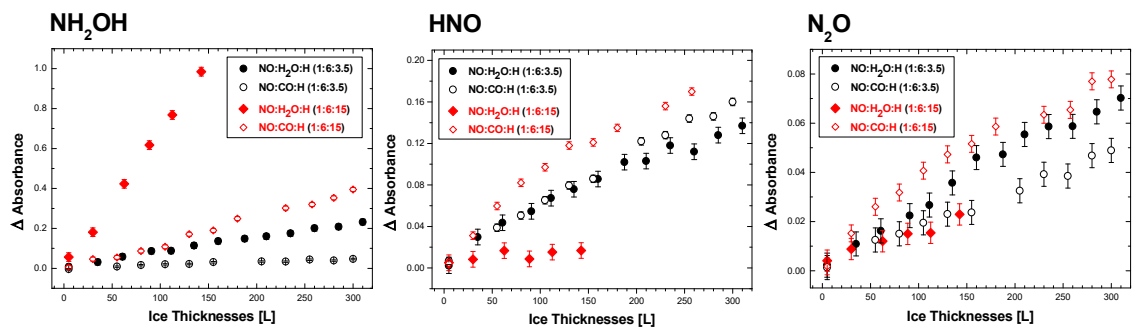

Figure 7. RAIR integrated intensities of hydroxylamine (left-side panel), nitroxyl (centre panel) and nitrous oxide (right-side panel) formed upon H-atom exposure versus the ice thickness in a co-deposition experiment of NO:CO(1:6) + H and $NO:H_2O$(1:6) + H. Each panel shows four curves: low H-atom flux co-deposited with $NO:H_2O$ = 1:6 mixture (full black circles); low H-atom flux co-deposited with NO:CO = 1:6 mixture (empty black circles); high H-atom flux co-deposited with $NO:H_2O$ = 1:6 mixture (full red diamonds); low H-atom flux co-deposited with NO:CO = 1:6 mixture (empty red diamonds).

The NHOH and $NH_2O$ intermediates, as well as final product $NH_2OH$, form strong hydrogen bonds with a water lattice. HNO also forms this kind of bond, but less efficiently. Such hydrogen bonds help HNO and $NH_2OH$ molecules to dissipate excess energy quickly upon formation and make the hydrogenation of NO and HNO kinetically favourable compared to the case of NO + H in a CO lattice in which these kinds of bonds are not formed. It should be noted that a water matrix significantly decreases the $N_2O$ final yield for high H-atom flux, but increases it for low flux. Moreover, hydrogenation of NO in an $N_2$ matrix ($NO:N_2$ = 1:5) for high H-atom flux gives an $N_2O$ yield that is three times higher than that obtained in an NO:CO hydrogenation experiment. The interpretation of this result is not straightforward and the influence of different lattice types on the $N_2O$ formation route is therefore uncertain.

### G. Possible reaction pathways

*1. NO hydrogenation network*

Figure 8 shows the NH$_2$OH formation reaction scheme as investigated in this work. The solid lines indicate reaction pathways that take place without an activation barrier or with a very small activation barrier and are therefore efficient at cryogenic temperatures, *e.g.*, in dense cold quiescent clouds. The dashed lines represent the reactions that proceed with an activation barrier.

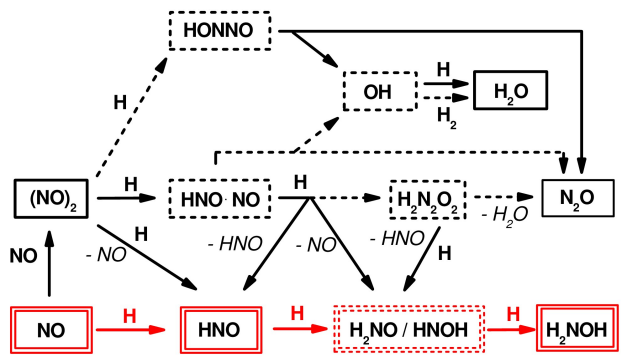

Figure 8. A schematic representation of the solid state NO + H reaction network, summarizing the conclusions from this work and Ref. 22. Solid squares indicate species that are detected in the ice while dashed squares indicate intermediate products that are not detected but may be formed in the ice. Solid arrows represent efficient and/or barrierless reactions. Dashed arrows represent reaction pathways with an activation barrier. Only atom and radical addition reactions are presented here (*e.g.*, hydrogen abstraction reactions are omitted for simplicity).

In Fig. 8, three subsequent H-atom addition reactions to form NH$_2$OH from a single NO molecule are highlighted in red double-bordered boxes.

$$\text{NO} + \text{H} \rightarrow \text{HNO} \tag{1a}$$
$$\text{HNO} + \text{H} \rightarrow \text{H}_2\text{NO} \tag{2a}$$
$$\text{H}_2\text{NO} + \text{H} \rightarrow \text{H}_2\text{NOH}. \tag{3a}$$

These three reactions represent the simplest and most efficient formation route of hydroxylamine. In reactions 1a and 2a, the H atoms are subsequently added to the N atom of the NO molecule, while the last H atom in reaction 3a is added to the oxygen. The reactions

$$\text{NO} + \text{H} \rightarrow \text{NOH} \tag{1b}$$
$$\text{HNO} + \text{H} \rightarrow \text{HNOH}, \tag{2b}$$

where the addition of H atoms takes place on the oxygen atom side of the NO molecule, are expected to be less efficient. Matrix isolation experiments[34] and *ab initio* calculations[52] indeed indicate that these reactions have high activation barriers (>4000 K and >10000 K, respectively)[52] and are, therefore, less efficient compared to the barrierless reactions 1a and 2a. However, for the reaction

$$\text{HNO} + \text{H} \rightarrow \text{NO} + \text{H}_2 \tag{2c}$$

gas-phase *ab initio* calculations[52] show that the rate of hydrogen atom abstraction is higher than the rate of hydrogen atom addition at all temperatures (*e.g.*, by a factor of 20 at 300 K). This is not consistent with our experimental results where hydroxylamine is formed even at very low H-atom fluxes. This discrepancy, however, is likely due to differences between solid-state and gas-phase reactions. Reactions 2a and 2c are both exothermic. Reaction 2c provides two final products and in this case the excess energy of the reaction can be efficiently distributed over two bodies to be translationally dissipated both in the gas phase and in the solid state. For reaction 2a this excess energy must be distributed only over the vibrational- and in the gas phase also rotational-levels of a single hydrogenation product (unless photon emission with required energy is allowed), and will be much less efficient in the gas phase than in the solid phase where the surrounding molecules play the role of an absorbing third body. Therefore, the presence of a third body can efficiently change the branching ratio between H-atom addition and H-atom subtraction reactions, leading, for instance, to the efficient surface formation of $NH_2OH$ ice. As in reaction 2c, the hydrogen abstraction reaction

$$H_2NO + H \rightarrow HNO + H_2 \quad (3b)$$

is also barrierless[51, 52] and should compete with the hydrogen addition reaction 3a. However, as discussed above, the formation of hydroxylamine has a higher effective branching ratio than reaction 3b in the solid state.

The hydrogenation of NO monomers through reactions 1a – 3a takes place only in $NO:CO$ and $NO:H_2O$ hydrogenation experiments where the CO or $H_2O$ lattice can keep a fraction of the NO monomers isolated and can prevent their dimerization through reactions 4a and 4b:

$$NO + NO \rightarrow cis\text{-}(NO)_2 \quad (4a)$$
$$NO + NO \rightarrow trans\text{-}(NO)_2. \quad (4b)$$

Formation of $cis$-$(NO)_2$ is the preferred reaction pathway and is confirmed by NO matrix isolation experiments.[28, 35] Under our experimental conditions $trans$-$(NO)_2$ is observed only in an $NO:CO:H$ = 1:6:70 co-deposition experiment where it is trapped by the CO lattice and the products of NO hydrogenation.

Due to a lack of gas-phase kinetic data the hydrogenation reaction scheme for $(NO)_2$ has so far been unclear. Here we experimentally show several reaction routes involving the hydrogenation of $(NO)_2$. First, the nitric oxide dimer reacts with an H atom to form HNO and NO,

$$(NO)_2 + H \rightarrow HNO + NO, \quad (5a)$$

or produces an HNO • NO associated complex. The existence of this complex has already been suggested by several groups,[53-56]

$$(NO)_2 + H \rightarrow HNO \cdot NO. \quad (5b)$$

The addition of another H atom to a nitrogen atom of this associated complex leads to the formation of $H_2NO$, HNO and NO:

$$HNO \cdot NO + H \rightarrow H_2NO + NO \quad (6a)$$
$$HNO \cdot NO + H \rightarrow HNO + HNO. \quad (6b)$$

All these reactions lead to the formation of hydroxylamine precursors as shown by reactions 1 – 3. Our experiments confirm that hydroxylamine will eventually be formed upon hydrogenation of these species.

There is another final product of NO hydrogenation reactions which is always observed in our experiments: $N_2O$ ice. The reaction pathway to form $N_2O$ is not yet clear, since all known possible mechanisms involve a high activation barrier.

A possible mechanism to form $N_2O$ is through HNO dimerization and the subsequent dissociation of the formed hyponitrous acid,[51, 53, 54]

$$HNO + HNO \rightarrow H_2N_2O_2 \text{ (OHN=NHO)} \quad (7)$$
$$H_2N_2O_2 \text{ (OHN=NHO)} \rightarrow \text{isomerisation} \rightarrow N_2O + H_2O. \quad (8)$$

This mechanism is rather unlikely to occur at cryogenic temperatures, because reaction 8 involves at least one H-atom migration from a nitrogen atom to an oxygen atom of an OHN=NHO molecule leading to the formation of the intermediate HON(NO)H or HON=NOH which will then dissociate to form nitrous oxide and water.[51, 53] Both dissociation and H-atom migration involve activation barriers as well as the HNO dimerization itself (reaction 7). However, several of these activation barriers can be overcome by the following mechanism. First, an H atom is added to an N atom of the HNO • NO associated complex. This reaction is exothermic and takes place without an activation barrier. The excess energy of this reaction can then help to overcome the $H_2N_2O_2$ dissociation barrier leading to the formation of nitrous acid by a mechanism similar to reactions 7 and 8:

$$HNO \cdot NO + H \rightarrow H_2N_2O_2(\text{"hot"}) \rightarrow \text{isomerisation} \rightarrow N_2O + H_2O. \quad (9)$$

The weak point of both suggested mechanisms is the high activation barrier in the last step of reactions 8 and 9. Because of this barrier $H_2N_2O_2$ (OHN=NHO) should be detectable in our spectra under our experimental conditions together with $N_2O$. Moreover, the infrared spectrum of $H_2N_2O_2$ and its deuterated analogue have strong absorption features, *i.e.*, two close bands around 1000 cm$^{-1}$ which are not present in our RAIR spectra (see Figs. 1 (c, e), Figs. 2 (a, c)).[57] The absence of $H_2N_2O_2$ absorption features indicates that another reaction pathway is likely responsible for the $N_2O$ formation.

One of these possible formation mechanisms for $N_2O$ was suggested earlier in Refs. 58, 59, where reduction of NO with alkaline hydroxylamine was studied:

$$HNOH + NO \rightarrow HON(NO)H \quad (10)$$

$$HON(NO)H \rightarrow N_2O + H_2O. \quad (11)$$

Reaction 10 is exothermic[51] and, as other radical-radical reactions, should proceed without an activation barrier. The product of reaction 10 may be formed with enough internal energy to overcome its dissociation barrier and to produce nitrous oxide and water (reaction 11). However, the formation of HNOH (reaction 2b) is thermodynamically unfavourable.[51] Therefore another formation mechanism is proposed here, *i.e.*, the direct addition of an H atom to the oxygen atom of an (NO)$_2$ dimer followyed the dissociation of the unstable radical:

$$ONNO + H \rightarrow (ONNOH) \rightarrow N_2O + OH. \quad (12)$$

This reaction pathway is exothermic, and according to BAC – MP4 calculations[51] the second step is barrierless (or at least has a very small barrier). The first step of this reaction is similar to reaction 1b and, therefore, presents an activation barrier (dashed arrow in Fig. 8). Although our experimental data are not conclusive on the efficiency of the $N_2O$ formation mechanisms, reaction 12 is likely the most efficient pathway because of its simplicity and because we do not detect the intermediates formed according to reactions 7 – 11.

In Ref. 22, two reaction pathways are suggested for the formation of $N_2O$ ice in a low surface coverage regime (<1 monolayer) on silicate surfaces. The first one is reaction 12 as discussed above. A different formation route is:

$$HNO + NO \rightarrow N_2O + OH. \quad (14a)$$

BAC-MP4 calculations[51] show that this formation mechanism involves at a certain stage the migration of one H atom from a nitrogen to an oxygen atom:

$$HNO + NO \rightarrow HN(O)NO \rightarrow ONNOH \rightarrow N_2O + OH. \quad (14b)$$

This migration process has a very high activation barrier (122 kJ mol$^{-1}$) and is endothermic.[51] A possible way to overcome this activation barrier under our experimental conditions is through the participation of a "hot" HNO molecule formed from reaction 1a before the HNO dissipates its excess energy (203 kJ mol$^{-1}$)[51] into the bulk of the ice. However this reaction is unlikely to occur in a high surface coverage regime, when HNO is always surrounded by other species. Moreover, reaction 14 requires at the same time the presence of NO monomer, which is detected in a high surface coverage regime only if NO is diluted into a CO matrix (see Fig 2a). Therefore, reaction 14 is unlikely to be efficient under our experimental conditions.

The unidentified absorption feature in our IR spectra, here named X-NO, appears at 1829 cm$^{-1}$. Further experimental constraints are needed to assign this feature to the HNO · NO complex or to the HON(NO)H intermediate.

The produced OH radicals from reaction 12 can react with H atoms and NO molecules to form:

$$OH + H \rightarrow H_2O \quad (13a)$$

$$OH + NO \rightarrow HNO_2. \quad (13b)$$

Recently another solid-state reaction pathway involving the OH radical was extensively investigated at cryogenic temperatures in Refs. 60, 61:

$$OH + H_2 \rightarrow H_2O + H. \tag{13c}$$

The identification of $H_2O$ and $HNO_2$ is not trivial due to the low final yield of these species and the overlap of their absorption features with other molecules present in the ice. Newly formed $H_2O$ molecules are reported in the gas phase upon desorption from the ice in Ref. 22. The amount of water ice formed at the end of the experiments described there is estimated to be a fraction of a monolayer (~0.5 ML). Moreover, all the suggested formation routes of $H_2O$ ice have $N_2O$ as an accompanying product. Since $N_2O$ is present only in low abundances in our experiments, the formation of $H_2O$ in the present study is expected to be inefficient.

As previously mentioned $H_2CO$ is not formed during low H-atom flux experiments, but it is present in the ice only if all NO is converted to its hydrogenation products. Both gas-phase experimental work[62] and calculations[51] show that the following reaction:

$$HCO + NO \rightarrow HNO + CO \tag{15}$$

is exothermic and likely barrierless. If reaction 15 takes place under our experimental conditions, the presence of NO in the ice can efficiently reduce the final $H_2CO$ yield in favour of the $NH_2OH$ formation. This is consistent with our experimental results. However, there is no direct evidence for this reaction pathway in our experiments. In particular, the low efficiency of $H_2CO$ formation may also be explained by a short residence time of the H atoms on the ice surface combined with the competition between the CO + H reaction, which has a barrier, and the barrierless NO + H reaction (1a).

*2. UV processing of the ice*

As opposed to the neutral-neutral non-energetic surface reactions discussed in the previous section, UV processing of ices provides an energetic input that leads to the dissociation of molecular bonds and the formation of "hot" fragments, *i.e.*, radicals, atoms and ions. These "hot" fragments can in turn react with the surrounding molecules and easily overcome activation barriers, desorb from the ice surface or diffuse into the bulk of the ice and kick other molecules out.[12, 26, 27] Here we focus on the UV photolysis of pure solid NO and NO mixed with CO and $H_2O$ ices. Although a detailed quantitative analysis of the UV-induced reaction network is challenging without knowing photodissociation and photoionisation cross-sections for all the molecules present and/or formed in the ice upon UV photolysis, we can still draw important conclusions on the efficiency of several reaction pathways.

UV processing of pure NO ice gives two final products: $N_2O$ and $N_2O_3$. These species can be formed through the following reaction:

$$2(NO)_2 + h\nu \rightarrow N_2O + N_2O_3. \tag{16}$$

First, a N-O bond from a *cis*-$(NO)_2$ is broken upon UV photolysis (reaction 17a). Then, the free oxygen atom recombines with a nearby nitrogen atom of another $(NO)_2$ dimer to form dinitrogen trioxide (reaction 18a):

$$(NO)_2 + h\nu \rightarrow N_2O + \cdot O \tag{17a}$$

$$(NO)_2 + \cdot O \rightarrow N_2O_3. \tag{18a}$$

This result is consistent with Ref. 63 where the photolysis of *cis*-$(NO)_2$ is studied at 13 K in an Ar matrix. The formation of nitrous oxide via direct photodissociation of nitric oxide and the consequent N-atom addition reaction to an NO dimer:

$$(NO)_2 + h\nu \rightarrow NO + \cdot N + \cdot O \tag{17b}$$

$$(NO)_2 + N \rightarrow N_2O + NO \tag{18b}$$

is unlikely to occur since gas-phase chemical data show[51, 53] that the nitrogen atom addition to NO forms molecular nitrogen rather than nitrous oxide. Moreover, the non-detection of the ·NCO radical in NO:CO photolysis experiments constrains the low efficiency of reaction 17b:

$$CO + N \rightarrow \cdot NCO. \tag{19}$$

UV processing of NO:CO ice mixtures yields two additional products: NO monomers and $CO_2$. The formation of NO monomers can be explained by the reaction:

$$(NO)_2 + h\nu \rightarrow NO + NO^* \tag{20}$$

and the subsequent isolation of "hot" monomers in a CO lattice.

The other product is $CO_2$ ice that can be formed through "hot" oxygen atom addition to carbon monoxide:

$$CO + \cdot O \rightarrow CO_2. \tag{21}$$

This reaction pathway is discussed in several experimental studies.[64] According to Ref. 64, reaction (21) takes place even at cryogenic temperatures, although it has a significant activation barrier. The transfer of O atoms from *cis*-$(NO)_2$ molecules is also observed in Ref. 63 after the exposure of Ar/NO/CO samples to photons at λ=220 – 330 nm.

Our experiments show that photolysis of nitric oxide in a matrix of carbon monoxide and water produces an even larger amount of $CO_2$. As recently shown in independent laboratory experiments,[8-10] $CO_2$ can be efficiently formed through the reactions:

$$H_2O + h\nu \rightarrow \cdot OH + \cdot H \tag{22}$$

$$CO + \cdot OH \rightarrow HOCO \rightarrow CO_2 + H, \tag{23}$$

where direct dissociation of the HOCO complex leads to the formation of $CO_2$ ice.

Photodissociation of water ice (reaction 22) can also explain the formation of the final products only detected when an NO:$H_2O$ ice was exposed to UV photons: $H_2O_2$ by recombination of two OH radicals, HNO by H-atom addition to NO (reaction 1a), and possibly $HNO_2$ by recombination of nitric oxide with a hydroxyl radical:

$$NO + \cdot OH \rightarrow HONO \tag{24}$$

Additionally, water can react with a "hot" oxygen atom to form hydrogen peroxide:

$$H_2O + \cdot O \rightarrow \cdot OH + \cdot OH \rightarrow H_2O_2 \tag{25}$$

This reaction pathway competes with the formation of $N_2O_3$ in water-rich matrices (reactions 17a - 18a) by consuming free O atoms and may be the cause of the low final yield of $N_2O_3$ detected in water-rich ices (see Fig. 4).

The newly formed products may themselves participate in UV induced chemistry. For example, UV photodissociation of $CO_2$ and $N_2O$ leads to the formation of free O atoms:

$$CO_2 + h\nu \rightarrow CO + \cdot O \tag{26}$$

$$N_2O + h\nu \rightarrow N_2 + \cdot O. \tag{27}$$

These reactions can play a role in oxygen transfer, but become important only after a long photolysis period with a total UV exposure of at least $10^{17}$ photons cm$^{-2}$. This flux corresponds to more than hundred incident UV photons per adsorption spot.

The low ionisation potential of *cis*-$(NO)_2$ can lead to the formation of $NO^+$ and $(NO)_2^+$ in the ice upon absorption of Ly-α radiation.[65] However, it is not clear from our experimental data to what extent ion chemistry is involved in the O atom transfer and the subsequent formation of $N_2O_3$ and $CO_2$.

The non-detection of $NH_2OH$ in the NO:$H_2O$ ice photolysis experiments indicates that there is a lack of free H atoms in the bulk of the ice. Therefore, reaction 22 has a low efficiency and/or recombination of hydrogen atoms is an efficient mechanism under our experimental conditions. Another mechanism which could reduce the final $NH_2OH$ yield is the efficient photodissociation of the intermediate HNO. Recently, we tried to form $NH_2OH$ by UV photolysis of $NH_3$:$H_2O$ and $NH_3$:$O_2$ ice mixtures. These experiments show that the formation of $NH_2OH$ in the bulk of $NH_3$:$H_2O$ ice is not efficient either, while UV photolysis of $NH_3$:$O_2$ ice mixtures gives uncertain results.

**IV. ASTROCHEMICAL IMPLICATIONS**

In quiescent dark clouds, grains provide surfaces on which species can accrete, meet and react and to which excess reaction energy can be donated. Grain surface chemistry is governed by the accretion rate of the gas-phase species and the surface migration rate. The timescale at which gas-phase species deplete onto grains is about $10^5$ years in dense clouds with lifetimes of dense cores between $10^5$ and $10^6$ years. Early during dense cloud formation an $H_2O$-rich (polar) ice containing mainly $CO_2$ and traces of $CH_4$ and $NH_3$ forms. Under these conditions NO can be also formed through surface reactions. In prestellar cores, where densities are as high as $10^5$ cm$^{-2}$ and temperatures are 10 – 20 K, gas-phase CO molecules freeze out onto the water ice layer forming a non-polar ice which is the site for additional and more complex reaction pathways. Under these conditions gas-phase formed NO can also deposit onto icy grains. Species like $N_2O$, HNO and NO itself can be formed through further surface reactions involving H-, N- and O-atoms. These species are, therefore, expected to be present in both polar and non-polar ices during the prestellar core-phase. In a recent astronomical study[21] we discussed the efficiency of the formation of hydroxylamine starting from the hydrogenation of a pure NO ice. In this work, by studying the hydrogenation of NO in $H_2O$-, CO-, and $N_2$- ice mantles, a full reaction scheme in polar and non-polar environments has been derived.

Due to the low abundance of interstellar solid NO compared to water ice the upper part of the chemical network presented in Fig. 8 is not likely to be astronomically relevant. The formation of an $(NO)_2$ dimer in space is indeed not favourable. However, the hydrogenation of NO molecules and the subsequent formation of $NH_2OH$ is efficient (through a barrierless mechanism) at cryogenic temperatures (15 K) in both water- and carbon monoxide-rich ices. Therefore, according to our experimental results we expect that in quiescent dark clouds NO is efficiently converted to hydroxylamine. However, special care is needed to extrapolate the laboratory data to astronomical timescales and fluxes, as our experiments, for obvious reasons, are not fully representative for interstellar conditions.

To show that the presented reaction scheme (Fig. 8) can indeed lead to the formation of $NH_2OH$ in dense cores, we have performed a gas-grain model using a full gas and grain chemical network of more than 6000 reactions between 460 species of which 195 are involved in both gas-phase and grain-surface reactions.[21] For this we used the OSU gas-grain code, first described by Ref. 66 with some recent modifications, which are described in Ref. 67. The initial abundances and chemical network are taken from a benchmark study of chemical codes.[68] The physical conditions are chosen similar to their "TMC 1" model which represents a prototypical dense cloud. A fraction of the formed surface species is allowed to desorb into the gas phase according to the Rice-Ramsperger-Kessel (RRK) theory[69, 70] with an efficiency of α = 0.01.

Most surface reactions in chemical networks are based on chemical intuition or are included by analogy to gas-phase reactions. The reaction between H and HNO in standard chemical networks leads to $H_2$ and NO, while $NH_2OH$ is usually formed through $NH_2$ + OH. Therefore, in standard chemical networks, no $NH_2OH$ is formed and NO ice is mostly converted into HNO or $N_2O$ under dark cloud conditions. This is in agreement with Garrod et al.[71] who performed chemical models of star formation and who found that $NH_2OH$ is only abundantly formed in the "hot-core" phase when the grain is lukewarm and $NH_2$ and OH become mobile.

Since in cold dense clouds the visual extinction is around 10 mag and photodissociation events are rare, the surface abundance of radical species like OH and $NH_2$ is low. The low grain-surface temperature of 10 K reduces the mobility of these species, while hydrogen atoms are already mobile at this temperature. Moreover, UV photolysis experiments show that hydroxylamine is not formed by irradiating $NO:CO:H_2O$ nor $NH_3:H_2O$ ice mixtures at low temperatures. The NO hydrogenation scheme as experimentally investigated in the present study is, therefore, much more promising as an $NH_2OH$ formation reaction route. In our model, that includes the NO hydrogenation scheme, $NH_2OH$ is indeed abundantly present in both the solid and the gas phase. The hydroxylamine abundance peaks at $10^5$ years. Photodissociation into OH and $NH_2$, of which the latter converts into $NH_3$ upon hydrogenation, starts to become important at this time. Snow *et al.*[72] showed that gas-phase $NH_2OH$ can react with $CH_5^+$ to form $NH_3OH^+$ that can in turn react with different carboxylic acids to form protonated amino acids. These reactions are not included in the networks used. Our model,[21] however, predicts that hydroxylamine has a relative abundance of at least $7 \cdot 10^{-9}$ $n_H$ on the grain surface and $3 \cdot 10^{-11}$ $n_H$ in the gas phase at the end of the lifetime of a dense cold cloud (1-10 Myr), before the core collapses.

To date, and to the best of our knowledge, interstellar hydroxylamine has not been observed in space, and only upper limits have been found for a number of sources.[73] These correspond to fractional abundances of $< 10^{-11}$ in the gas phase. This is around the same order of magnitude as our gas-phase model predictions for dense clouds. We therefore expect that in the near future the faint radio signal from $NH_2OH$ gas-phase molecules will be recorded by the receivers of ALMA that will investigate the inner parts of proto-planetary disks and hot-cores with relatively high temperatures.[74] The hydroxylamine locked onto the grains at the beginning of the core collapse becomes accessible for further reaction at later stages when a protostar forms and UV irradiation and thermal processing start. The astrobiological implications of our study are constrained by the results of Ref. 15. They proposed a route for the synthesis of amino acids in interstellar environments starting from desorption of $NH_2OH$. In their study $NH_2OH$ is formed through energetic processing at a late stage of star formation. Here we have shown how $NH_2OH$ is already formed in the solid phase though a non-energetic route in the early stages of star formation. Figure 9 shows the link between the updated solid-phase $NH_2OH$ formation scheme and the gas-phase formation routes of amino acids in space. Here the key stage of this suggested mechanism is that in which reactions between protonated hydroxylamine and carboxylic acids occur. Although these reaction pathways were already proved experimentally to be efficient at room temperature,[15] recent calculations show that they may involve activation barriers[75] and, therefore, that they may not be efficient under astronomically relevant conditions. Thus, there is now great demand for experimental verification of the latter results.

Finally, acetic acid, another main component of this reaction network, has already been detected towards several astronomical objects (mainly hot-cores)[76, 77] and Bennett at al.[78] showed successfully the laboratory synthesis of acetic acid by UV photolysis of $CO_2:CH_4$ ices under UHV conditions and cryogenic temperatures.

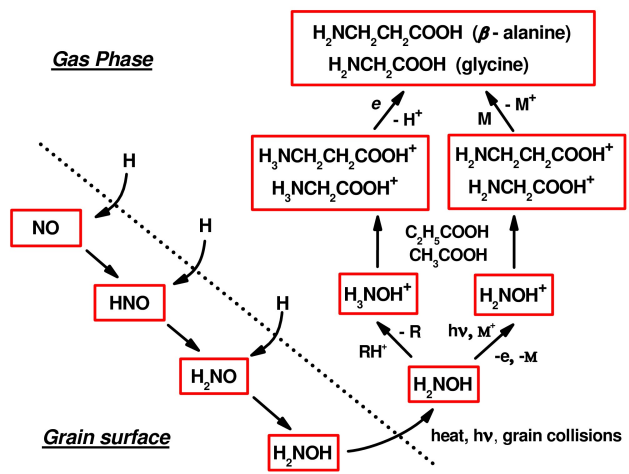

Figure 9. Proposed formation route of amino acids in star-forming regions based on the hydroxylamine surface formation route investigated in the present study combined with the experimental work of Ref. 15.

## V. CONCLUSIONS

A systematic study of NO hydrogenation pathways under UHV conditions and for cryogenic temperatures is presented. In particular, the influence of sample temperature, NO deposition rate, H-atom flux and UV photolysis on the formation of different hydrogenation products is studied. Special attention is given to the hydrogenation of NO in CO- and $H_2O$-rich ices and the astronomical implications of our results are discussed. The main conclusions of this work are:

1. In CO- and $H_2O$-rich interstellar ice analogues NO is efficiently converted to hydroxylamine upon exposure to thermal H atoms.
2. The formation of hydroxylamine takes place through a fast and barrierless mechanism by subsequent addition of three H atoms to a single NO molecule.
3. HNO is an intermediate product of hydroxylamine formation and is observed in our experiments only under low H atom flux conditions.
4. The final HNO and $NH_2OH$ formation yield is higher at 15 K than at 25 K.
5. In water-rich interstellar ice analogues the final $NH_2OH$ formation yield is higher than in CO-rich ices.
6. In CO-rich ices hydroxylamine is formed well before CO molecules are hydrogenated to form formaldehyde and methanol.
7. $N_2O$ is a side product of the NO hydrogenation reaction scheme.
8. Solid $NH_2OH$ is not formed upon UV (Ly-α) photolysis of $NO:CO:H_2O$ ices. The end products are $N_2O$, $N_2O_3$ (or $NO_2$), HNO, $H_2O_2$ and $CO_2$ ice.
9. According to our laboratory and modelling results we expect that $NH_2OH$ is already present in the solid phase in dark molecular clouds and mainly in the gas phase in protostar regions. We therefore expect that in the near future gas-phase $NH_2OH$ will be detected by ALMA.
10. Hydroxylamine formed efficiently in the solid phase provides a gas-phase reservoir upon desorption as a starting point in the formation of glycine and β-alanine.

## ACKNOWLEDGMENTS

We acknowledge many stimulating discussions with our French collaborators, who are authors on the accompanying paper. We furthermore thank the European Community's Seventh Framework Programme (FP7/2007-2013) under grant agreement no. 238258, the Netherlands Research School for Astronomy (NOVA), Netherlands Organization for Scientific Research (NWO) through a VICI grant and the European Research Council (ERC-2010-StG, Grant Agreement No. 259510-KISLMOL). We also would like to acknowledge Joseph Guss (Leiden Observatory) for his helpful comments.

## REFERENCES

[1] E. Herbst and E. F. van Dishoeck, Complex organic interstellar molecules. *Annu. Rev. Astro. Astrophys.* **47**, 427-480 (2009).

[2] V. Wakelam, I. V. M. Smith, E. Herbst, J. Troe, W. Geppert, H. Linnartz, K. Öberg, E. Roueff, M. Agúndez, P. Pernot, H. M. Cuppen, J. C. Loison, D. Talbi, Reaction networks for interstellar chemical modelling: improvements and challenges. *Space Sci. Rev.* **156**, 13-72 (2010).

[3] N. Watanabe, A. Kouchi, Efficient formation of formaldehyde and methanol by the addition of hydrogen atoms to CO in $H_2O$-CO ice at 10 K. Astrophys. J. **571**, 173-176 (2002), and references cited therein.

[4] G. W. Fuchs, H. M. Cuppen, S. Ioppolo, C. Romanzin, S. E. Bisschop, S. Andersson, E. F. van Dishoeck, H. Linnartz, Hydrogenation reactions in interstellar CO ice analogues. A combined

experimental/theoretical approach. *Astron. Astrophys.* **505,** 629-639 (2009).

[5] N. Miyauchi, H. Hidaka, T. Chigai, A. Nagaoka, N. Watanabe, A. Kouchi, Formation of hydrogen peroxide and water from the reaction of cold hydrogen atoms with solid oxygen at 10 K. *Chem. Phys. Lett.* **456,** 27-30 (2008).

[6] S. Ioppolo, H. M. Cuppen, C. Romanzin, E. F. van Dishoeck, H. Linnartz, Laboratory evidence for efficient water formation in interstellar ices. *Astrophys, J.* **686,** 1474-1479 (2008).

[7] E. Matar, E. Congiu, F. Dulieu, A. Momeni, J. L. Lemaire, Mobility of D atoms on porous amorphous water ice surfaces under interstellar conditions. *Astron. Astrophys.* **492,** L17-L20 (2008).

[8] Y. Oba, N. Watanabe, A. Kouchi, T. Hama, V. Pirronello, Experimental study of $CO_2$ formation by surface reactions of non-energetic OH radicals with CO molecules. *Astrophys. J.* **712,** L174-L178 (2010).

[9] S. Ioppolo, Y. van Boheemen, H. M. Cuppen, E. F. van Dishoeck, H. Linnartz, Surface formation of $CO_2$ ice at low temperatures,. *Mon. Not. R. Astron. Soc.* **413,** 2281-2287 (2011).

[10] J. A. Noble, F. Dulieu, E. Congiu, H. J. Fraser, $CO_2$ formation in quiescent clouds: an experimental study of the CO + OH pathway. *Astrophys. J.* **735,** 121 (2011).

[11] S. A. Sandford *et al.*, Organics captured from comet 81P/Wild 2 by the Stardust spacecraft. *Science* **314,** 1720-1724 (2006).

[12] K. I. Öberg, S. Bottinelli, J. K. Jørgensen, E. F. van Dishoeck, A cold complex chemistry toward the low-mass protostar B1-b: evidence for complex molecule production in ices, *Astrophys. J.* **716**, 825–834 (2010).

[13] G. M. Muñoz Caro, U. J. Meierhenrich, W. A. Schutte, B. Barbier, A. Arcones Segovia, H. Rosenbauer, W. H.-P. Thiemann, A. Brack, J. M. Greenberg, Amino acids from ultraviolet irradiation of interstellar ice analogues, Nature **416**, 403-406 (2002).

[14] S. B. Charnley, S. D. Rodgers, P. Ehrenfreund, Gas-grain chemical models of star-forming molecular clouds as constrained by ISO and SWAS observations. *Astron. Astrophys.* **378,** 1024-1036 (2001).

[15] V. Blagojevic, S. Petrie, D. K. Bohme, Gas-phase syntheses for interstellar carboxylic and amino acids. *Mon. Not. R. Astron. Soc.* **339,** L7-L11 (2003).

[16] H. S. Liszt, B. E. Turner, Microwave detection of interstellar NO. *Astrophys. J.* **224,** L73-L76 (1978).

[17] T. H. Pwa and S. R. Pottasch, Astron. Astrophys. 164, 116 (1986).

[18] D. McGonagle, L. M. Ziurys, W. M. Irvine, Y. C. Minh, Detection of nitric oxide in the dark clouds L134N. *Astrophys. J.* **359,** 121-124 (1990).

[19] M. Gerin, Y. Viala, F. Pauzat, Y. Ellinger, The abundance of nitric oxide in molecular clouds. *Astron. Astrophys.* **266,** 463-478 (1992).

[20] E. Herbst, W. Klemperer, The formation and depletion of molecules in dense interstellar clouds. *Astrophys. J.* **185,** 505-533 (1973).

[21] E. Congiu, G. Fedoseev, S. Ioppolo, F. Dulieu, H. Chaabouni, S. Baouche, J. L. Lemaire, C. Laffon, P. Parent, T. Lamberts, H. M. Cuppen, H. Linnartz. NO ice hydrogenation - a solid pathway to $NH_2OH$ formation in space. *Astrophys. J. Lett.* **750**, L12 (2012).

[22] E. Congiu *et al*. Efficient Surface Formation Route of Interstellar Hydroxylamine through NO Hydrogenation I: the submonolayer regime on interstellar relevant substrates. submitted to *J. Chem. Phys.* (2012).

[23] K. G. Tschersich, Intensity of a source of atomic hydrogen based on a hot capillary. *J. Appl. Phys.* **87**, 2565-2573 (2000).

[24] S. Ioppolo, H. M. Cuppen, C. Romanzin, E. F. van Dishoeck, H. Linnartz, Water formation at low temperatures by surface $O_2$ hydrogenation I: characterization of ice penetration. *PCCP* **12**, 12065-12076 (2010).

[25] H. M. Cuppen, S. Ioppolo, C. Romanzin, H. Linnartz, Water formation at low temperatures by surface $O_2$ hydrogenation II: the reaction network. PCCP **12**, 12077-12088 (2010).

[26] K. I. Öberg, R. T. Garrod, E. F. van Dishoeck, H. Linnartz, Formation rates of complex organics in UV irradiated $CH_3OH$-rich ices. I. Experiments. *Astron. Astrophys.* **504**, 891–913 (2009).

[27] K. I. Öberg, G. W. Fuchs, Z. Awad, H. J. Fraser, S. Schlemmer, E. F. van Dishoeck, H. Linnartz, Photodesorption of CO Ice. *Astrophys. J.* **662**, L23-L26 (2007).

[28] W. G. Fateley, H. A. Bent, B. Crawford, Infrared spectra of the frozen oxides of nitrogen. *J. Chem. Phys.* **31**, 204-217 (1959).

[29] E. M. Noir, L. H. Chen, M. M. Strube, J. Laane, Raman spectra and force constants for the nitric oxide dimer and its isotopic species. *J. Phys. Chem.* **88**, 756-759 (1984).

[30] R. Withnall, L. Andrews, Matrix infrared spectra and normal-coordinate analysis of isotopic hydroxylamine. *J. Phys. Chem.* **92**, 2155-2161 (1988).

[31] G. A. Yeo, T. A. Ford, The infrared spectrum of the hydroxylamine dimer. *J. Molec. Struct.* **217**, 307-323 (1990).

[32] R. E. Nightingale and E. L. Wagner, The vibrational spectra and structure of solid hydroxylamine and deutero-hydroxylamine. *J. Chem. Phys.* **22**, 203-208 (1954).

[33] C. S. Jamieson, C. J. Bennett, A. M. Mebel, R. I. Kaiser, Investigating the mechanism for the formation of nitrous oxide [$N_2O(X^1\Sigma^+)$] in extraterrestrial ices. *Astrophys. J.* **624**, 436-447 (2005).

[34] M. E. Jacox and D. E. Milligan, Matrix-isolation study of the reaction of H atoms with NO: The infrared spectrum of HNO. *J. Molec. Spectrosc.* **48**, 536-559 (1973).

[35] E. J. Sluyts and B. J. Van der Veken, On the behavior of nitrogen oxides in liquefied argon and krypton. Dimerisation of nitric oxide. *J. Molec. Struct.* **320**, 249-267 (1994).

[36] W. J. Dulmage, E. A. Meyers, W. N. Lipscomb, On the crystal and molecular structure of $N_2O_2$. *Acta Cryst.* **6**, 760-764 (1953).

[37] C. S. Barrett and L. Meyer, Molecular Packing, Defects, and Transformations in Solid Oxygen. *Phys. Rev.*, **160**, 694-697 (1967).

[38] D. T. Cromer, D. Schiferl, R. LeSar, R. L. Mills, Room-temperature structure of carbon monoxide at 2.7 and 3.6 GPa. *Acta Cryst.* **C39**, 1146–1150 (1983), and references cited therein.

[39] J. E. Bertie and M. R. Shehata, The infrared spectra of $NH_3 \cdot H_2O$ and $ND_3 \cdot D_2O$ at 100 K. *J. Chem. Phys.* **83**, 1449-1456 (1985).

[40] E. M. Noir, L. H. Chen, J. Laane, Interconversion studies and characterization of asymmetric and symmetric dinitrogen trioxide in nitric oxide matrices by raman and infrared spectroscopy. *J. Phys. Chem.* **87**, 1113-1120 (1983).

[41] M. H. Moore and R. L. Hudson, Infrared study of ion-irradiated $N_2$ – dominated ices relevant to Triton and Pluto: formation of HCN and HNC. *Icarus*, **161**, 486-500 (2003).

[42] M. Pettersson, L. Khriachtchev, S. Jolkkonen, M. Räsänen, Photochemistry of HNCO in solid Xe: channels of UV photolysis and creation of $H_2NCO$. *J. Phys. Chem.* A. **103**, 9154-9162 (1999).

[43] V. E. Bondybey, J. H. English, C. Weldon Mathews, R. J. Contolini, Infrared spectra and isomerisation of CHNO species in rare gas matrices. *J. Molec. Spectrosc.* **92**, 431-442 (1982).

[44] R. L. Hudson, R. K. Khanna, M. H. Moore, Laboratory evidence for solid-phase protonation of HNCO in interstellar ices. *Astrophys. J. Suppl. Ser.* **159**, 277-281 (2005).

[45] A. Yabushita, T. Hama, D. Iida, N. Kawanaka, M. Kawasaki, N. Watanabe, M. N. R. Ashfold, H.-P. Loock, Release of hydrogen molecules from the photodissociation of amorphous solid water and polycrystalline ice at 157 and 193 nm. *J. Chem. Phys.* **129**, 044501 (2008).

[46] J. B. Bossa, K. Isokoski, M. de Valois, H. Linnartz, submitted to *Astron. Astrophys.* (2012).

[47] H. M. Cuppen and E. Herbst, Simulation of the formation and morphology of ice mantles on interstellar

grains. *Astrophys. J.* **668**, 294-309 (2007).

[48] C. P. Smyth and K. B. McAlpine, The dipole moment of nitric oxide. *J. Chem. Phys*. **1**, 60-61 (1933), and references cited within.

[49] W. T. Rawlins, J. C. Person, M. E. Fraser, S. M. Miller, W. A. M. Blumberg, The dipole moment and infrared transition strengths of nitric oxide. *J. Chem. Phys*. **109**, 3409-3417 (1998).

[50] C. M. Western, P. R. R. Langridge-Smith, B. J. Howard, The rotational spectrum and structure of the nitric oxide dimer. OSU international symposium on molecular spectroscopy (1980).

[51] E. W. G. Diau, M. C. Lin, Y. He, C. F. Melius, Theoretical aspects of H/N/O-chemistry relevant to the thermal reduction of NO by $H_2$. *Prog. Energy Combust. Sci.* **21**, 1-23 (1995).

[52] M. Page and M. R. Soto, Radical addition to HNO. *Ab initio* reaction path and variation transition state theory calculations for H+HNO→$H_2$NO and H+HNO→HNOH. *J. Chem. Phys.* **99**, 7709-7717 (1993).

[53] M. C. Lin, Y. He, C. F. Melius, Theoretical interpretation of the kinetics and mechanisms of the HNO + HNO and HNO + 2NO reactions with a unified model. *Int. J. Chem. Kin.* **24**, 489-516 (1992).

[54] S. G. Cheskis, V. A. Nadtochenko, O. M. Sarkisov, Study of the HNO + HNO and HNO + NO reactions by intracavity laser spectroscopy. *Int. J. Chem. Kin.* **13**, 1041 - 1050 (1981).

[55] W. A. Seddon, J. W. Fletcher, F. C. Sopchyshyn, Pulse radiolysis of nitric oxide in aqueous solution. *Can. J. Chem.* **51**, 1123-1130 (1973).

[56] D. Chakraborty, C.-C. Hsu, M. C. Lin, Theoretical studies of nitroamino radical reactions: Rate constants for the unimolecular decomposition of $HNNO_2$ and related bimolecular processes. *J. Chem. Phys.* **109**, 8887-8896 (1998).

[57] G. E. McGraw, D. L. Bernitt, I. C. Hisatsune, Infrared spectra of isotopic hyponitrite ions. *Spectrochim. Acta*, **23A**, 25-34 (1967).

[58] J. N. Cooper, J. E. Chilton, R. E. Powell, Reaction of nitric oxide with alkaline hydroxylamine. *Inorg. Chem.* **9**, 2303-2304 (1970).

[59] F. T. Bonner, L. S. Dzelzkalns, J. A. Bonucci, Properties of nitroxyl as intermediate in the nitric oxide-hydroxylamine reaction and trioxodinitrate decomposition. *Inorg. Chem.* **17**. 2487-2494 (1978).

[60] C. Romanzin, S. Ioppolo, H. M. Cuppen, E. F. van Dishoeck, H. Linnartz, Water formation by surface $O_3$ hydrogenation. *J. Chem. Phys.* **134**, 084504 (2011).

[61] Y. Oba, N. Watanabe, T. Hama, K. Kuwahata, H. Hidaka, and A. Kouchi, Water formation through a quantum tunneling surface reaction, OH + $H_2$, at 10 K, *Astrophys. J.* **749,** 67 (2012).

[62] N. I. Butkovskaya, A. A. Muravyov, D. W. Setser, Infrared chemiluminescence from the NO + HCO reaction: observation of the $2v_1$-$v_1$ hot band of HNO. *Chem. Phys. Lett.* **266**, 223-226 (1997).

[63] M. Hawkins and A. J. Downs, Photochemistry of argon matrices containing nitric oxide and carbonyl sulfide. 1. The photolysis of the nitric oxide dimer, cis-$N_2O_2$, *J. Phys. Chem*. **88**, 1527-1533 (1984).

[64] U. Raut and R. A. Baragiola, Solid-state CO oxidation by atomic O: a route to solid $CO_2$ synthesis in dense molecular clouds. *Astrophys. J. Lett.* **737**, L 14 (2011), and references cited within.

[65] S. V. Levchenko, H. Reiser, A. I. Krylov, O. Gessner, A. Stolow, H. Shi, A. L. L. East, Photodissociation dynamics of the NO dimer. I. Theoretical overview of the ultraviolet singlet excited states, *J. Chem. Phys.* **125**, 084301 (2006).

[66] T. I. Hasegawa, E. Herbst, C. M. Leung, Models of gas-grain chemistry in dense interstellar clouds with complex organic molecules. *Astrophys. J. Supp. Ser.* **82,** 167-195 (1992).

[67] G. E. Hassel, E. Herbst, R. T. Garrod, Modeling the lukewarm corino phase: is L1527 unique? *Astrophys. J.* **681,** 1385-1395 (2008).

[68] D. Semenov, F. Hersant, V. Wakelam, A. Dutrey, E. Chapillon, S. Guilloteau, T. Henning, R. Launhardt, V. Pietu, K. Schreyer, Chemistry in disks. IV. Benchmarking gas-grain chemical models with surface reactions. *Astron. Astrophys.* **522,** A42 (2010).

[69] K. A. Holbrook, M. J. Pilling, S. H. Robertson, *Unimolecular reactions.* (Wiley, Second Edition, New York, 1996).

[70] R. T. Garrod, V. Wakelam, E. Herbst, Non-thermal desorption from interstellar dust grains via exothermic surface reactions. *Astron. Astrophys.* **467,** 1103-1115 (2007).

[71] R. T. Garrod, S. L. Widicus Weaver, E. Herbst, Complex chemistry in star-forming regions: an expanded gas-grain warm-up chemical model. *Astrophys. J.* **682,** 283-302 (2008).

[72] J. L. Snow, G. Orlova, V. Blagojevic, D. K. Bohme, Gas-phase ionic synthesis of amino acids: beta versus alpha. *J. Am. Chem. Soc*. **129,** 9910-9917 (2007).

[73] R. L. Pulliam, B. A. McGuire, A. J. Remijan, A search for hydroxylamine ($NH_2OH$) towards selected astronomical sources. *Astrophys. J.* **751:1** (2012).

[74] E. Herbst, Chemistry in the ISM: the ALMA (r)evolution. The cloudy crystal ball of one astrochemist. *Astrophys. Space Sci.* **313,** 129-134 (2008).

[75] C. Barrientos, P. Redondo, L. Largo, V. M. Rayon, A. Largo, Gas-phase synthesis of precursors of interstellar glycine: a computational study of the reactions of acetic acid with hydroxylamine and its ionized and protonated derivatives, *Astrophys. J.* **748**, 99 (2012).

[76] D. M. Mehringer, L. E. Snyder, Y. Miao, F. J. Lovas, Detection and confirmation of interstellar acetic acid. *Astrophys. J.* **480,** L71-L74 (1997).

[77] A. Remijan, L. E. Snyder, D. N. Friedel, S.-Y. Liu, R. Y. Shah, A survey of acetic acid towards hot molecular cores. *Astrophys. J.* **590,** 314-332 (2003).

[78] C. J. Bennett and R. I. Kaiser, The formation of acetic acid ($CH_3COOH$) in interstellar ice analogs. *Astrophys. J.* **660,** 1289-1295 (2007).